\documentclass[%
twocolumn, 
english,
amsmath,amssymb,
aps,
pra,
 floatfix,
]{revtex4-1}

\usepackage{graphicx}
\usepackage{dcolumn}
\usepackage{bm}
\usepackage{physics}
\usepackage[usenames, dvipsnames]{color}
\usepackage{sidecap}
\usepackage{textcomp}

\begin{document}
\title{Atomic-scale imaging of a 27-nuclear-spin cluster \\ using a single-spin quantum sensor}

\author{M. H. Abobeih$^{1,2}$}
\author{J. Randall$^{1,2}$}
\author{C. E. Bradley$^{1,2}$}
\author{H. P. Bartling$^{1,2}$}
\author{M. A. Bakker$^{1,2}$}
\author{M. J. Degen$^{1,2}$}
\author{M. Markham$^3$} 
\author{D. J. Twitchen$^3$}
\author{T. H. Taminiau$^{1,2}$}
 \email{T.H.Taminiau@TUDelft.nl}

\affiliation{$^{1}$QuTech, Delft University of Technology, PO Box 5046, 2600 GA Delft, The Netherlands}
\affiliation{$^{2}$Kavli Institute of Nanoscience Delft, Delft University of Technology, PO Box 5046, 2600 GA Delft, The Netherlands}
\affiliation{$^{3}$Element Six, Fermi Avenue, Harwell Oxford, Didcot, Oxfordshire, OX11 0QR, United Kingdom}

\date{\today}



\maketitle


\textbf{Nuclear magnetic resonance (NMR) is a powerful method for determining the structure of molecules and proteins \cite{rule2006fundamentals}. While conventional NMR requires averaging over large ensembles, recent progress with single-spin quantum sensors \cite{mamin2013nanoscale, staudacher2013nuclear, shi2015single, Lovchinsky_Science2016, Aslam_Science2017, glenn2018high, smits2019two,lovchinsky2017magnetic} has created the prospect of magnetic imaging of individual molecules \cite{ajoy2015atomic, kost2015resolving, perunicic2016quantum, wang2016positioning}. As an initial step towards this goal, isolated nuclear spins and spin pairs have been mapped \cite{sushkov2014magnetic, Muller_NatComm2014, Shi_NatPhys2014, Zopes_NatComm2018, Zopes_PRL2018,sasaki_prb2018, Abobeih_NatComm2018, Yang_arXiv2019}. However, large clusters of interacting spins ---  such as found in molecules ---  result in highly complex spectra. Imaging these complex systems is an outstanding challenge due to the required high spectral resolution and efficient spatial reconstruction with sub-angstrom precision. Here we develop such atomic-scale imaging using a single nitrogen-vacancy (NV) centre as a quantum sensor, and demonstrate it on a model system of $27$ coupled $^{13}$C nuclear spins in a diamond. We present a new multidimensional spectroscopy method that isolates individual nuclear-nuclear spin interactions with high spectral resolution ($< 80\,$mHz) and high accuracy ($2$ mHz). We show that these interactions encode the composition and inter-connectivity of the cluster, and develop methods to extract the 3D structure of the cluster with sub-angstrom resolution. Our results demonstrate a key capability towards magnetic imaging of individual molecules and other complex spin systems \cite{ajoy2015atomic, kost2015resolving, perunicic2016quantum, wang2016positioning,lovchinsky2017magnetic}.}


The nitrogen-vacancy (NV) centre in diamond has emerged as a powerful quantum sensor \cite{mamin2013nanoscale, staudacher2013nuclear, Lovchinsky_Science2016, Aslam_Science2017, glenn2018high,lovchinsky2017magnetic, shi2015single, smits2019two,ajoy2015atomic,kost2015resolving,perunicic2016quantum, wang2016positioning,rosenfeld2018sensing,knowles2016demonstration}. The NV electron spin provides long coherence times \cite{Lovchinsky_Science2016,Aslam_Science2017,Abobeih_NatComm2018} and high-contrast optical readout \cite{Lovchinsky_Science2016, Cramer_NatureComm2016,pfender2019high}, enabling high sensitivity over a large range of temperatures  \cite{Lovchinsky_Science2016,Aslam_Science2017,Abobeih_NatComm2018,pfender2019high,Cujia_arXiv2018}. Pioneering experiments with near-surface NV centers have demonstrated spectroscopy of small ensembles of nuclear spins in nano-scale volumes \cite{mamin2013nanoscale, staudacher2013nuclear, smits2019two,Lovchinsky_Science2016, Aslam_Science2017, glenn2018high}, and electron-spin labelled proteins \cite{shi2015single}. Furthermore, single nuclear spin sensitivity has been demonstrated and isolated individual nuclear spins and spin pairs have been mapped \cite{Muller_NatComm2014,sushkov2014magnetic, Zopes_NatComm2018, Zopes_PRL2018,sasaki_prb2018, Shi_NatPhys2014, Abobeih_NatComm2018, Yang_arXiv2019}. Together, these results have established the NV center as a promising platform for magnetic imaging of complex spin systems and single molecules \cite{ajoy2015atomic,kost2015resolving,perunicic2016quantum, wang2016positioning}. 

In this work, we realise a key ability towards that goal: the 3D imaging of large nuclear-spin structures with atomic resolution. The main idea of our method is to obtain structural information by accessing the couplings between individual nuclear spins. The key open challenges are: (1) to realize high spectral resolution so that small couplings can be accessed, (2) to isolate such couplings from complex spectra, and (3) to transform the revealed connectivity into the 3D spatial structure with sub-angstrom precision.

\begin{figure*}
\includegraphics[scale=0.85]{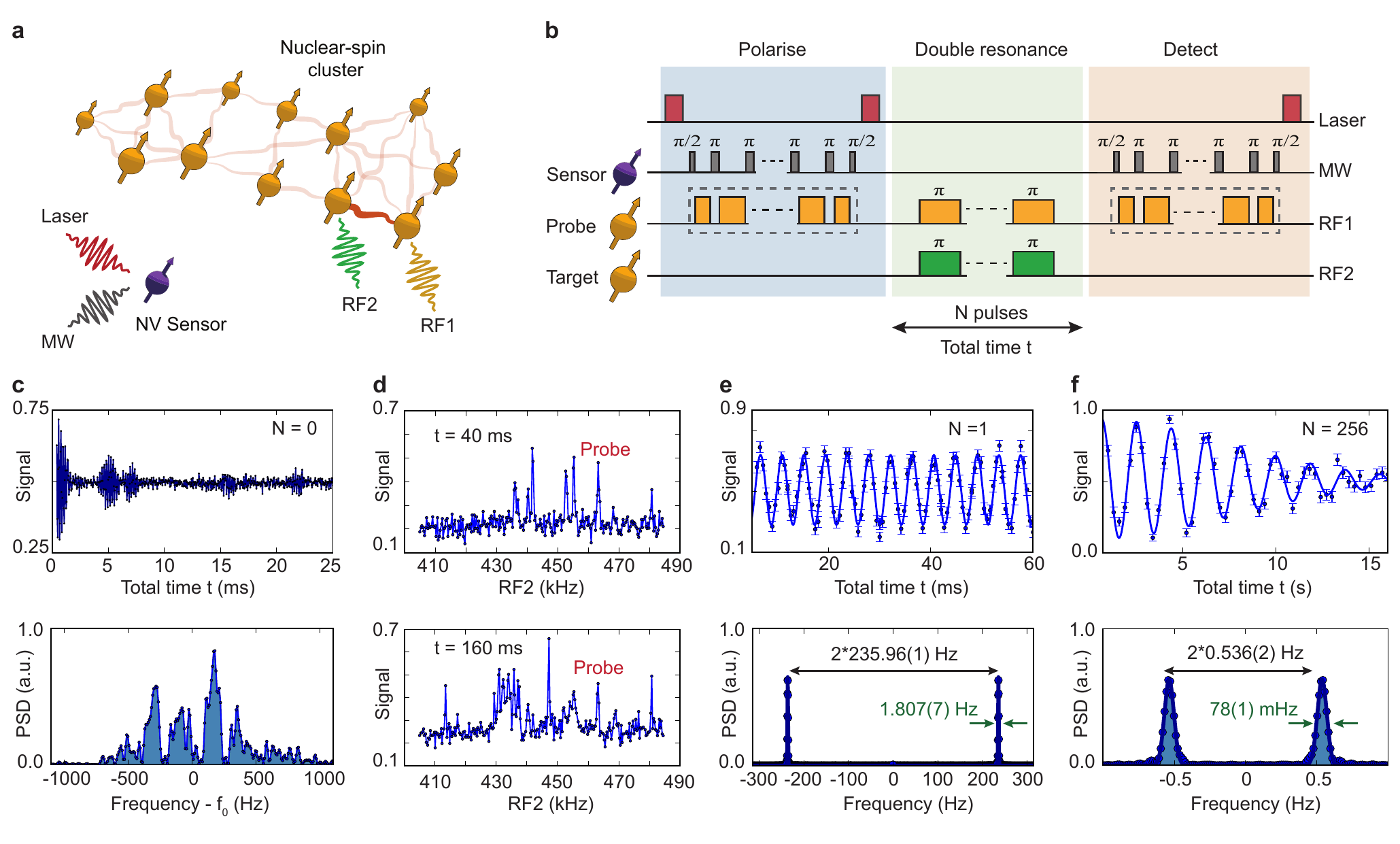}
\caption{\label{Figure1}
\textbf{Basic concepts of the experiment.} \textbf{a,} We consider an individual cluster of $^{13}$C nuclear spins near a single NV centre in diamond. To obtain the 3D structure of the cluster we use the NV electron spin as a quantum sensor to measure nuclear-nuclear spin couplings. \textbf{b,} Experimental sequence. The NV sensor is used to polarise and detect the ``probe'' spin(s) at frequency $RF1$ (see methods). A double-resonance sequence of $N$ echo pulses is applied simultaneously on the probe spin(s) ($RF1$) and the ``target'' spin(s) ($RF2$), so that the coupling between these spins is selectively detected. \textbf{c,} A Ramsey signal ($N=0$) for a nuclear spin in the cluster (detuning $f_0=5$ kHz). Because all couplings are probed simultaneously, the power spectral density (PSD) yields a complex non-resolvable spectrum. See Supplementary Fig. 1 for more examples. \textbf{d,} Double-resonance spectroscopy ($N=1$). Sweeping the target frequency ($RF2$) reveals all spins that couple to the probe spin(s). For larger $t$, more peaks appear as weaker couplings become visible. \textbf{e,} Sweeping the evolution time $t$ for a fixed $RF1$ and $RF2$ reveals the coupling strength between spins. This example reveals a $235.96(1)$ Hz coupling between two spins with a spectral resolution of 1.807(7) Hz FWHM. \textbf{f,} An example with $N=256$ echo pulses showing an extended coherence time to $10.6(6)$ seconds, which enables selective measurements of sub-Hz couplings with high spectral resolution ($78(1)$ mHz) and precision ($2$ mHz). All graphs: see methods for fit functions.}
\end{figure*}

The basic elements of our experiment are illustrated in Fig. \ref{Figure1}a. We consider a cluster of $^{13}$C nuclear spins in the vicinity of a single NV centre in diamond at $4$ K. This cluster provides a model system for the magnetic imaging of single molecules and spin structures external to the diamond. Each $^{13}$C spin precesses at a shifted frequency due to the hyperfine interaction with the electron spin, resembling a chemical shift in traditional NMR \cite{Slichter_1996,rule2006fundamentals}. These shifts enable different nuclear spins in the cluster to be distinguished.

We use the NV electron spin as a sensor to probe the nuclear-nuclear interactions (Fig. 1b). Inspired by NMR spectroscopy \cite{Slichter_1996,rule2006fundamentals}, we develop sequences that employ spin-echo double-resonance (SEDOR) techniques to isolate and measure individual couplings with high spectral resolution. First, we polarise a nuclear ``probe" spin (frequency $RF1$) using newly developed quantum sensing sequences that can detect spins in any direction from the NV, enabling access to a large number of spins (see methods) \cite{Bradley_arXiv2019}. Second, we let this probe spin evolve for a time $t$ and apply $N$ echo pulses that decouple it from the other spins and environmental noise. Simultaneously, pulses on a ``target" spin in the cluster (frequency $RF2$) re-couple it to the probe spin, selecting the interaction between these two spins. Finally, a second sensing sequence detects the resulting polarisation of the probe spin through a high-contrast readout of the electron spin (see methods) \cite{Cramer_NatureComm2016}, which enables fast data collection. This double-resonance sequence provides a high spectral resolution through a long nuclear phase accumulation time. Importantly, the resolution is not limited by the relatively short coherence time of the electron spin sensor (see methods) \cite{Cramer_NatureComm2016, laraoui2013high}.\\


It is instructive to first consider the case without echo pulses ($N=0$), for which all couplings act simultaneously. This results in complex spectra that indicate many nuclear-nuclear spin interactions and/or simultaneous signals from multiple spins (Fig. \ref{Figure1}c). The  underlying structure of individual spins and couplings is obscured by the many frequencies ($2^j$ for coupling to $j$ spins) and the low spectral resolution of $\sim\,$30 Hz FWHM, set by the nuclear spin linewidth.

The echo sequence enables couplings between specific spins to be isolated and measured with high resolution. We first scan the target frequency $RF2$ for a fixed probe frequency $RF1$ (Fig. \ref{Figure1}d). This reveals the spectral positions of  nuclear spins coupled to the probe spin. We then sweep the evolution time $t$ and Fourier transform the signal to quantify the coupling strengths (Fig. \ref{Figure1}e). For a single echo pulse ($N = 1$), the nuclear spin coherence time is $0.56(2)\,$s, yielding a spectral resolution of $1.807(7)\,$Hz and a centre frequency accuracy of $10$ mHz. The spectral resolution can be further enhanced by applying more echo pulses. For $N=256$, a resolution of $78(1)$ mHz and an accuracy of $2$ mHz are obtained for the spin in Fig. 1f, making it possible to detect sub-Hertz interactions. The obtained resolution is an order of magnitude higher than in previous experiments on spin signals  \cite{maurer2012room,pfender2019high,smits2019two,Pfender_NatComm2017, glenn2018high,Aslam_Science2017}.

To characterise the complete cluster, we perform 3D spectroscopy by varying the probe frequency $RF1$, the target frequency $RF2$, and the evolution time $t$. The combinations of $RF1$ and $RF2$ reveal the spectral positions of the spins in the cluster. The coupling between spins is retrieved from the Fourier transform of the time dimension $t$. This yields a 3D data set that in principle encodes the composition and connectivity of the spin cluster (Fig. 2).    

\begin{figure}
\includegraphics[scale=0.88]{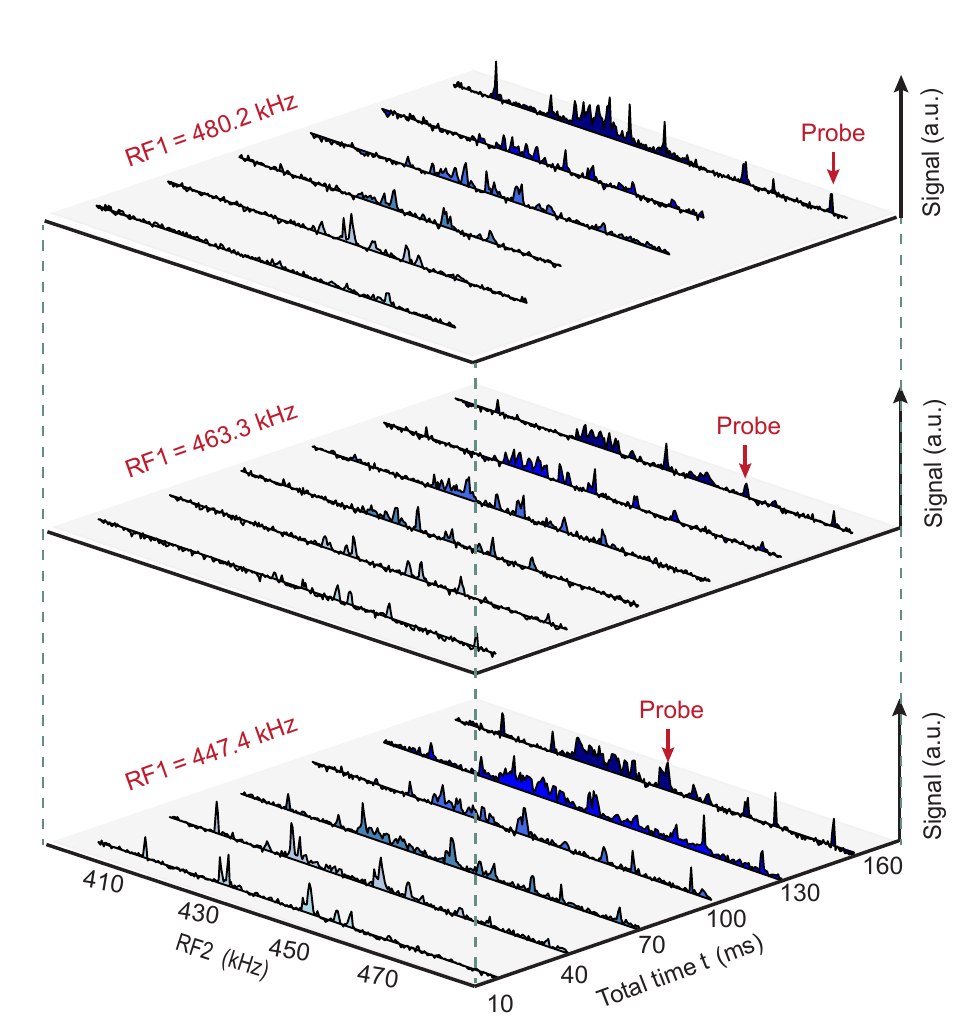}
\caption{\label{Figure2}
\textbf{Three-dimensional spectroscopy.} By varying the probe frequency $RF1$, the target frequency $RF2$, and the evolution time $t$, we obtain a three-dimensional data set that encodes the composition of the spins in the cluster and their couplings. The observation of a signal at $\{RF1,RF2\}$ indicates the presence of one or more spins at both frequencies and a coupling between them. Fourier transforming the time dimension $t$ reveals the spin-spin coupling strength. Examples for three different $RF1$ values are shown.}
\end{figure}

\begin{figure}
\includegraphics[scale=0.85]{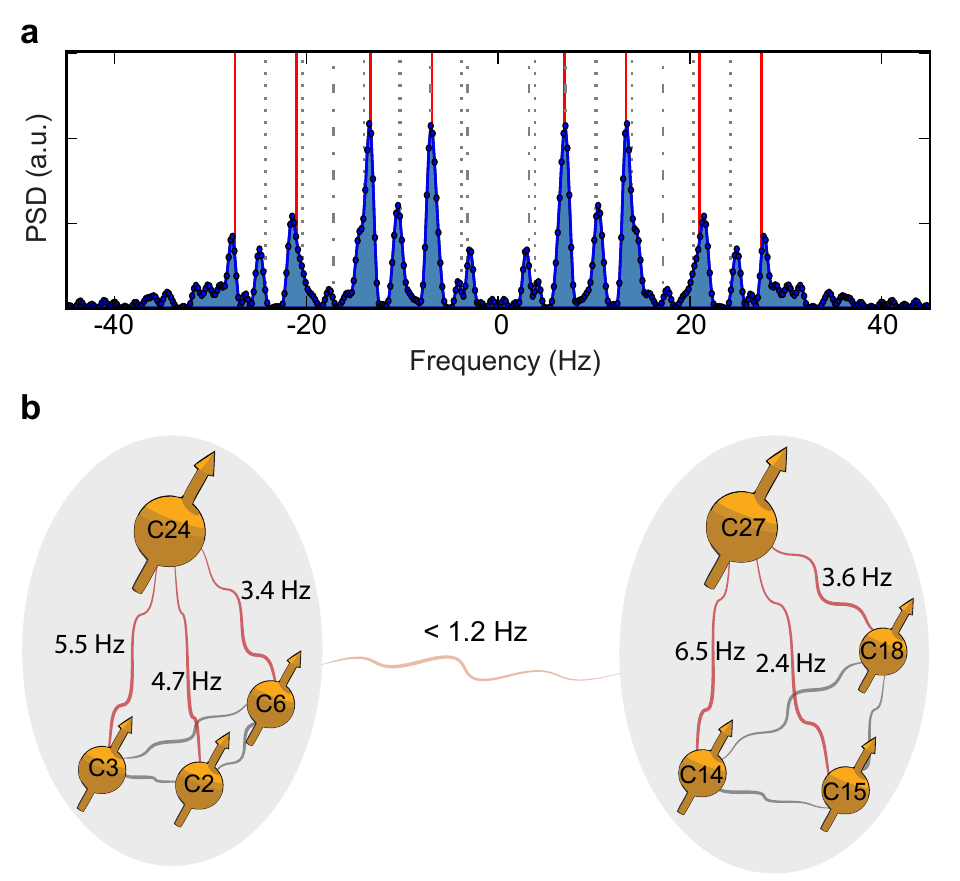}
\caption{\label{Figure3}
\textbf{Resolving ambiguities due to spectrally overlapping spins.} \textbf{a,} Retrieving couplings when multiple spins act simultaneously. Example in which the pulses invert three target spins (quadruple resonance). The PSD reveals a complex, yet resolvable, spectrum. Red lines indicate the frequencies $f = \pm f_1 \pm f_2 \pm f_3$, where $f_1 = 17.17(2)$ Hz, $f_2 = 7.05(3)$ Hz and $f_3 = 3.21(4)$ Hz are the extracted couplings of the probe spin to three target spins. Grey dashed lines mark additional frequency components that appear due to failures to invert one or two of the target spins (see Supplementary Fig. 2 for detailed analysis). \textbf{b,} Identifying spins and assigning couplings. Example from our data. Spins C2, C3, C6, C14, C15 and C18 all yield a coupling signal to the same RF2 frequency. Because these $6$ probe spins are part of two spatially separated clusters it follows that the signal at RF2 must originate from two distinct target spins (C24 and C27).}  
\end{figure}


To identify the spins and their couplings from the 3D spectra, we need to resolve ambiguities due to overlapping signals from multiple spins at near-identical frequencies. The first challenge is to retrieve individual couplings when multiple couplings are probed at the same time. Figure 3a shows an example where the couplings between one probe spin and three target spins are measured simultaneously. Whilst the spectrum becomes more complex, the high spectral resolution of our method enables retrieval of the underlying couplings. The second challenge is to determine the number of spins in the cluster and to assign the measured couplings to them. While the observation of a coupling at frequencies $RF1$ and $RF2$ is by itself not enough to assign it to particular spins, our method extracts multiple couplings that together constrain the problem. In particular, each spin couples predominantly to spins in its vicinity, so that spins can be uniquely identified from their connections to the rest of the cluster (see Fig. 3b for an example). 


\begin{figure*}
\includegraphics[scale=0.85]{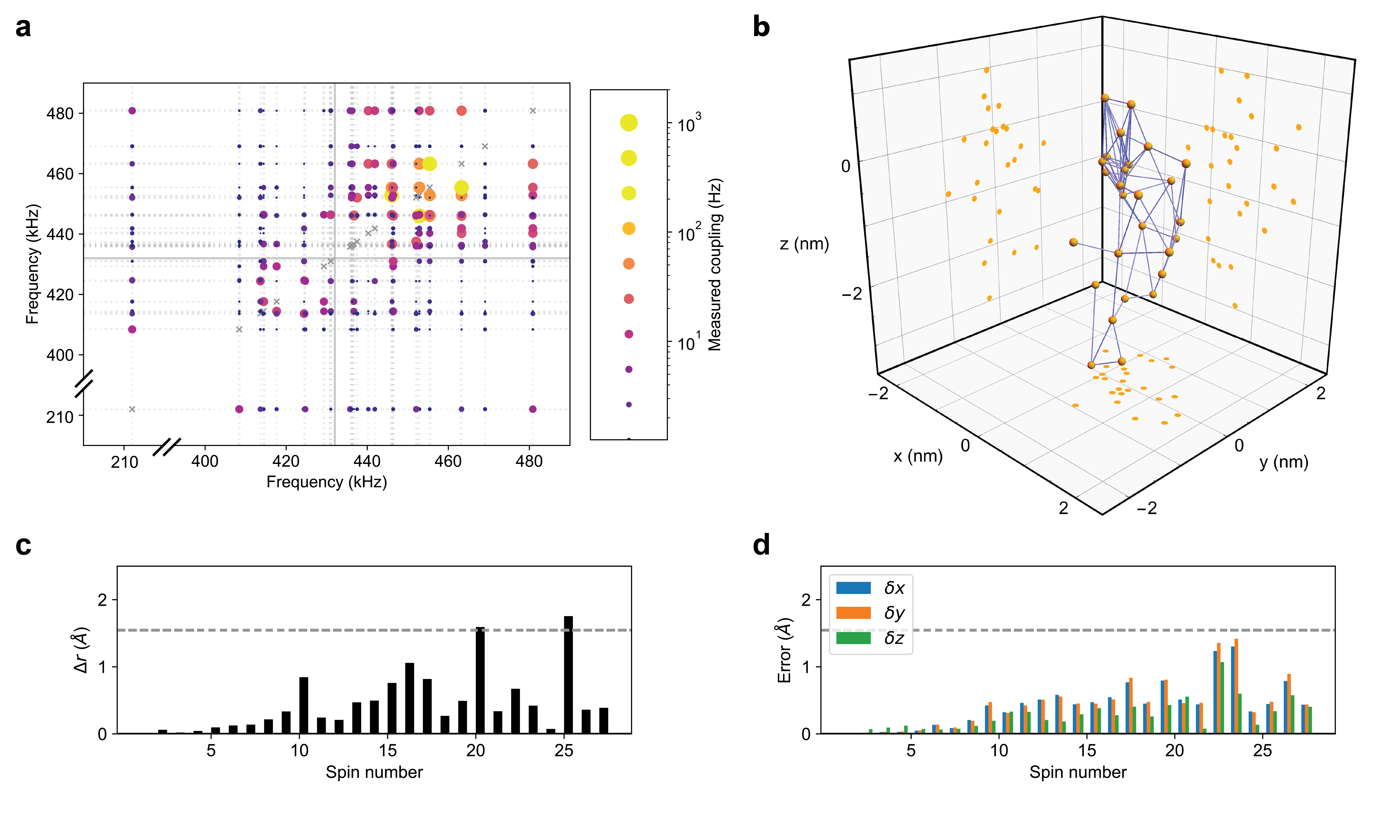}
\caption{\label{Figure4}
\textbf{Atomic-scale imaging of the 27-nuclear-spin cluster.}  \textbf{a,} 2D plot summarising all couplings between the 27 spins identified from the 3D spectroscopy (Fig. 2). Includes identification of spins with overlapping frequencies following Fig. 3. The size and colour of each point indicates the strength of the measured coupling averaged over the electron $m_s=+1$ and $m_s=-1$ states. Dashed grey lines indicate the nuclear spin frequencies ($m_s = -1$ state). Solid grey lines indicate the bare $^{13}$C Larmor frequency. Total measurement time: $\sim 400$ hours. See Supplementary Tables 2-4 for numerical values and uncertainties. \textbf{b,} 3D structure of the nuclear spins obtained using the diamond-lattice method (see text). Blue lines indicate couplings greater than $3\,$Hz and illustrate the connectivity of the cluster. \textbf{c,} Distance $\Delta r$ between the obtained spin positions from the diamond-lattice method (see text) and from a least-squares optimisation. Deviations are generally below one diamond bond length (dashed line, $\sim 1.54\,$\AA). \textbf{d,} The uncertainties for the 77 spatial coordinates of the cluster from a least-squares optimisation are less than the bond length, indicating atomic-scale resolution. See Supplementary Figs 6, 8 and 10 and Supplementary Table 5 for in depth comparisons between the structures and uncertainties obtained with the different methods.}
\end{figure*}

Transforming the 3D spectra into a spatial structure requires a precise relation between the measured couplings and the relative positions of the spins. A complication is that the presence of electronic spins can modify the nuclear couplings \cite{dutt2007quantum}, causing the measured value to deviate from a basic dipole-dipole coupling. We use perturbation theory to derive a set of many-body corrections that depend on the electron-nuclear and nuclear-nuclear couplings, and the magnetic field direction (see methods). For the type of cluster considered here, the corrections could be significant. However, the signs of the leading terms depend on the electron spin state. By averaging the measured couplings for the $m_s = +1$ and $m_s = -1$ states, the deviations are strongly reduced. Together with a novel method to align the magnetic field to within $0.07$ degrees (see methods), this enables us to approximate the nuclear-nuclear couplings as dipolar.


Finally, we determine the structure of the spin cluster. Figure \ref{Figure4}a summarises all extracted couplings. We identify $M=27$ nuclear spins and retrieve a total of $171$ pairwise couplings, out of the total of $M(M-1)/2 = 351$ couplings. The structure of the cluster is completely described by $3M-4 = 77$ spatial coordinates (see methods), so that the problem is overdetermined. However, due to the large number of parameters and local minima, a direct least-squares minimisation \cite{ajoy2015atomic} is challenging. Instead, we sequentially build the structure by progressively adding spins, while keeping track of all possible structures that match the measured couplings within a certain tolerance.

We use two different methods. The first method constrains the spin coordinates to the diamond lattice. The second method discretises space in a general cubic lattice, with voxel spacing down to $5\times10^{-3}\,$nm ($\sim 1/70$th of the lattice constant, see methods). While this second method is more computationally intensive, it uses minimum a priori knowledge and can be applied on arbitrary spin systems. We run these analyses in parallel with the measurements, so that sets of the most promising spin assignments and structures are regularly created. These yield predictions for which unmeasured couplings (combinations of $RF1$ and $RF2$) are required to decide between different assignments and structures, which we use to guide the experiments and reduce the total measurement time (see methods). 
 
Figure \ref{Figure4}b shows the structure obtained for the $27$ spins using the diamond-lattice. The blue connections show the strongest couplings ($>3$ Hz) and visualise the inter-connectivity of the cluster. The cubic-lattice method yields a nearly identical structure (see methods); the average distance between the spin positions for the two solutions is $0.58\,$\AA, a fraction of the bond length of $\sim 1.54\,$\AA. As a final step, we use these structures as inputs for least-squares minimisation, where the $x,y,z$ coordinates are allowed to relax to any value. The solution obtained lies close to the initial guess with an average distance of $0.46$ \AA. The uncertainties for the spatial coordinates ($\delta x$, $\delta y$, $\delta z$) are below a diamond bond length for all 27 spins (Fig. \ref{Figure4}c,d), indicating atomic-scale imaging of the complete $27$-spin cluster.

Additionally, we determine the position of the NV sensor relative to the cluster. Although not required to reconstruct the cluster, this provides a control experiment. We measure the coupling of the $^{14}$N nuclear spin to $12$ of the $^{13}$C spins (Supplementary Fig. 9). This unambiguously determines the location of both the $^{14}$N atom and the vacancy (fit uncertainties $< 0.3\,$\AA). We can now compare the electron-$^{13}$C hyperfine couplings to previous density functional theory (DFT) calculations for $5$ of our spins \cite{Nizovtsev_2018}. All $5$ couplings agree with the DFT calculations (Supplementary Fig. 9), providing an independent corroboration of the extracted structure, as well as a direct test of the DFT calculations. Looking beyond quantum sensing, this precise microscopic characterisation of the NV environment provides new opportunities for improved control of quantum bits for quantum information \cite{Bradley_arXiv2019,Abobeih_NatComm2018,Cramer_NatureComm2016,maurer2012room,dutt2007quantum}, and for investigating many-body physics in coupled spin systems. 


In conclusion, we have developed and demonstrated 3D atomic-scale imaging of large clusters of nuclear spins using a single-spin quantum sensor. Our approach is compatible with room temperature operation \cite{maurer2012room,Pfender_NatComm2017,Cujia_arXiv2018,pfender2019high} and can be extended to larger structures, as the number of required measurements scales linearly with the number of spins. Future improvements in the data acquisition and the computation of 3D structures can further reduce time requirements. In particular, recent methods to polarise and measure nuclear spins are expected to improve sensitivity \cite{Cujia_arXiv2018,pfender2019high}, especially for samples with weak couplings to the NV sensor. Optimised sampling of the measurements \cite{Yang_arXiv2019} and adaptive algorithms based on a real-time structure analysis can further reduce the total number of required measurements. Therefore, when combined with recent progress in nanoscale NMR with near-surface NV centres \cite{mamin2013nanoscale, staudacher2013nuclear, Lovchinsky_Science2016, Aslam_Science2017, glenn2018high, shi2015single, smits2019two}, our results provide a path towards the magnetic imaging of individual molecules and complex spin structures external to diamond \cite{ajoy2015atomic, kost2015resolving, perunicic2016quantum, wang2016positioning}. \\



\noindent\textbf{METHODS}

\noindent\textbf{Sample and NV centre sensor.} We use a naturally occurring NV centre in a homoepitaxially chemical-vapor-deposition (CVD) grown diamond with a $1.1\%$ natural abundance of $^{13}$C and a $\langle 111 \rangle$ crystal orientation (Element Six). The NV is placed in a solid-immersion lens to enhance photon collection efficiency \cite{Robledo_Nature2011}. The NV centre has been selected for the absence of $^{13}$C spins with hyperfine couplings $>$ 500 kHz. The NV electron spin coherence times are $T_2^* = 4.9(2)\,\mu$s and $T_2 = 1.182(5)\,$ms.

The NV sensor spin is used to create and detect polarisation (Fig. 1b). The two key requirements for the sensor spin are (1) a high-contrast readout to keep measurement times manageable, and (2) that it does not limit the spectral resolution by disturbing the phase evolution of the nuclear spins through relaxation \cite{maurer2012room,Pfender_NatComm2017}. We work at 4 Kelvin, so that the electron relaxation is negligible ($T_1=3.6(3) \times 10^3\,$s \cite{Abobeih_NatComm2018}), and use high-fidelity readout through resonant optical excitation (average $F = 94.5\%$) \cite{Robledo_Nature2011}. Note that recent experiments have satisfied both these requirements at room temperature \cite{Lovchinsky_Science2016,maurer2012room,Pfender_NatComm2017,pfender2019high,Cujia_arXiv2018}, so that the spectroscopy and imaging methods developed here can be applied at ambient conditions. 


\noindent\textbf{Magnetic field alignment.} 
A magnetic field of $\sim403\,$G is applied using a permanent magnet. We align the magnetic field along the NV axis to avoid electron-mediated shifts that cause the measured couplings to deviate from nuclear-nuclear dipolar coupling (see Supplementary Information section III). We use a ``thermal'' echo sequence --- previously introduced to measure temperature \cite{Toyli8417} --- to decouple the electron spin from magnetic noise along the NV axis, while retaining the sensitivity to the magnetic field in the $x,y$ directions (see Supplementary Fig. 4). This extends the sensing time from $T_2^* \approx 5\,\mu$s to $T_2 \approx 1$ ms, resulting in an uncertainty in the alignment of $0.07$ degrees.

\noindent\textbf{Quantum sensing sequences.} 
We employ two different sensing sequences. Sequence A consists of dynamical decoupling sequences of $N$ equally spaced $\pi$-pulses on the electron spin of the form ($\tau -\pi - \tau)^N$ \cite{Taminiau_PRL2012,Kolkowitz_PRL2012,Zhao_NatureNano2012}. This sequence is only sensitive to nuclear spins with a significant electron-nuclear hyperfine component perpendicular to the applied magnetic field \cite{Taminiau_PRL2012}. The inter-pulse spacing 2$\tau$ determines the spin frequency that is being probed. 

Sequence B is a newly developed method, described in more detail in Bradley et al. \cite{Bradley_arXiv2019}, that interleaves the dynamical decoupling sequence with RF pulses. This method enables the detection of spins with a weak or negligible perpendicular hyperfine component \cite{Bradley_arXiv2019,Pfender_NatComm2017}, so that spins in any direction from the NV can be detected. In this work, this enables us to access a greater number of spins in the cluster. For this sequence, the frequency of the RF pulse sets the targeted spin frequency, while $\tau$ can be freely chosen \cite{Bradley_arXiv2019}. 

\noindent\textbf{Electron-nuclear spectroscopy.}
As a starting point, we use the electron spin as a sensor to roughly characterise some of the nuclear spins in the cluster. We perform spectroscopy by sweeping the interpulse delay $\tau$ in sequence A (see for example Abobeih et al. \cite{Abobeih_NatComm2018}) and the RF frequency for sequence B \cite{Bradley_arXiv2019}. This identifies the frequency range at which spins are present in the cluster and provides the parameters to polarise and detect several spins \cite{Cramer_NatureComm2016}. 

\noindent\textbf{Nuclear-nuclear double-resonance spectroscopy.} 
The sequence for the double resonance experiments is given in Fig. \ref{Figure1}b. To polarise and detect the probe spin, we either use sequence A (without the RF1 pulses in the dashed box) or sequence B (with the RF1 pulses), depending if the perpendicular hyperfine coupling to the electron spin is significant or not. For sequence A, we set the interpulse delay as $\tau = (2k-1)\pi/(\omega_0 + RF1)$, with $k$ an integer and $\omega_0$ the $^{13}$C Larmor frequency for the electron $m_s=0$ state, and calibrate the number of pulses $N$ to maximise the signal \cite{Taminiau_PRL2012}. For sequence B we calibrate the RF power to maximise the signal. 

We create nuclear polarisation by projective measurements \cite{Cramer_NatureComm2016}. First the electron is prepared in a superposition state through resonant excitation \cite{Robledo_Nature2011} and a $\pi/2$ pulse. Second, the sensing sequence correlates the phase of the electron with the nuclear spin state. Finally, the electron is read out so that the nuclear spin is projected into a polarised state \cite{Cramer_NatureComm2016}. To enhance the signal-to-noise ratio and to ensure that the electron measurement does not disturb the nuclear spin evolution, we only perform the double resonance sequence if a photon was detected during the electron readout \cite{Cramer_NatureComm2016}. The resulting signal contrast for different spins varies from $20\%$ to $96\%$.

For the double resonance sequence, the phases of the RF1 echo pulses  are calibrated so that the phase difference with respect to the polarisation axis is $0$ or $\pi/2$. For the target spins, the phase of the RF2 pulse does not affect the signal and is arbitrarily set.

To mitigate pulse errors we alternate the phases of the pulses following the XY8 scheme \cite{XY8}, both for the electron and nuclear spins. For the electron spin, we use Hermite pulse envelopes \cite{warren1984effects} with Rabi frequency $\sim 14\,$MHz to obtain effective microwave pulses without initialisation of the intrinsic $^{14}$N nuclear spin. The nuclear-spin Rabi frequencies are in the range $0.3- 0.7\,$kHz.

\noindent\textbf{Data analysis.} 
We extract the spin-spin couplings $f$ and their uncertainties from fitting the time-domain double resonance signals (e.g. Fig. 1e-f, top) to $S  = a + A\cdot e^{-(t/T_2)^2}\cos{(2\pi{f} t+ \phi)}$, where $T_2$ is the coherence time (also a fit parameter). The PSD is obtained from a Fourier transform of the time domain signal with zero filling \cite{rule2006fundamentals} and the D.C. component filtered out (e.g. Fig. 1e-f, bottom). The spectral resolution (FWHM) is obtained from a Gaussian fit of the PSD, and its uncertainty is obtained from the fit of the time domain signal. Alternatively we can define the spectral resolution (FWHM) directly from the time domain signal as $\frac{2\sqrt{ln 2}}{\pi T_2}$, which yields $0.945(7)$ Hz for Fig. 1e and $50(1)$ mHz for Fig. 1f. 

\noindent\textbf{Electron-mediated interactions.}
We calculate corrections to the nuclear-nuclear couplings using perturbation theory up to second order. In contrast to previous results for strong electron-nuclear couplings \cite{dutt2007quantum,Zhao_NatureNat2011}, here many-body interactions due to the non-secular nuclear-nuclear couplings must be taken into account. The resulting frequency in a double resonance experiment is of the form (see Supplementary Information section IV)

\begin{equation}\begin{split}
    f_\text{DE}&(m_s = \pm 1) \approx \frac{1}{4\pi}|C\ + \\
    &+\Delta\lambda_1(m_s) + \Delta\lambda_2(m_s) + \Delta\lambda_3(m_s) |,
\end{split}\end{equation}
where $C$ is the parallel ($zz$) component of the dipole-dipole interaction between the nuclear spins and $\Delta\lambda_i$ are correction terms due to the presence of the electron spin. See Supplementary Information for the full analysis of all terms.

The dominant correction for our parameter regime is $\Delta\lambda_2$, which depends on both the electron-nuclear and nuclear-nuclear interactions. We make a Taylor expansion up to first order in $A_{zz}^{(j)}/\gamma_c B_z$, where $A_{zz}^{(j)}$ is the parallel electron-nuclear hyperfine coupling for spin $j$, $\gamma_c$ is the nuclear gyromagnetic ratio and $B_z$ is the component of the magnetic field along the NV axis. This yields an expression of the form $\Delta\lambda_2(m_s) \approx m_s \Delta\lambda_2^{(0)} + \Delta\lambda_2^{(1)}$, where the leading, zeroth-order, correction $m_s\Delta\lambda_2^{(0)}$ is given by

\begin{equation}
    \Delta\lambda_2^{(0)} = \frac{(A_{zx}^{(1)} + A_{zx}^{(2)}) C_{zx} + (A_{zy}^{(1)} + A_{zy}^{(2)}) C_{zy}}{\gamma_c B_z},
\end{equation}
where $A_{zx}^{(j)}$ ($C_{zx}$) and and $A_{zy}^{(j)}$ ($C_{zy}$) are the perpendicular electron-nuclear (nuclear-nuclear) coupling components. We cancel this term by averaging the double resonance frequencies measured for the $m_s = \pm 1$ electron spin projections. 

The remaining electron-mediated corrections depend on the angles of the electron-nuclear hyperfine interactions. Because these angles are unknown, we estimate the maximum possible shift for each spin-spin interaction by maximising over all angles. For our cluster (Fig. \ref{Figure4}), most of these maximum possible shifts are small (their average value is $\sim 0.03\,$Hz). In rare cases, the maximum possible correction runs up to $0.6\,$Hz (see Supplementary Information section IV), but as the locations of the involved spins are already precisely fixed through strong ($>20\,$Hz) interactions with several other spins, this would have a negligible effect on the obtained structure. Therefore, we can base the structural analysis on dipole-dipole interactions.

\noindent\textbf{3D structure analysis.}
The 3D structure of the nuclear spins is obtained using the dipole-dipole coupling formula, which relates the $zz$ couplings $C_{ij}$ to the spatial $x,y,z$ coordinates of spins $i$ and $j$ as

\begin{equation}\label{eq:hyperfine}
    C_{ij} = \frac{\alpha_{ij}}{\Delta r_{ij}^3}\left(\frac{3(z_j - z_i)^2}{\Delta r_{ij}^2}-1\right),
\end{equation}
where $\Delta r_{ij} = \sqrt{ (x_j - x_i)^2 + (y_j - y_i)^2 + (z_j - z_i)^2 }$, $\alpha_{ij} = \mu_0 \gamma_i \gamma_j \hbar/4\pi$, $\mu_0$ is the permeability of free space, $\gamma_i$ is the gyromagnetic ratio of nuclear spin $i$ and $\hbar$ is the reduced Planck constant. 

The goal is to minimise the sum of squares $\xi = \sum_{i<j}|\Delta f_{ij}|^2$, where $\Delta f_{ij} = f_{ij} - |C_{ij}|/4\pi$ are the residuals and $f_{ij}$ are the measured coupling frequencies. For $M = 27$ spins, there are $3M - 4 = 77$ free coordinates and $M(M-1)/2 = 351$ pairwise couplings, of which $171$ were determined in this work. $\xi$ can in principle be minimised using standard fitting methods, however tests with randomly generated spin clusters indicate that the initial guess for the coordinates should be within $\sim0.5\,$\AA\ in order for the fit to converge to the correct solution. For 27 spins, this corresponds to an intractable $\sim 10^{100}$ possible initial guesses. Instead we sequentially build the structure by adding spins one-by-one. 

For the diamond lattice positioning method, we first use the strongest measured coupling to any spin that is already positioned to reduce the position of a new spin to a number of possible lattice coordinates. For each possible coordinate, we then check if the predicted couplings to all other spins satisfy $\Delta f_{ij} < \mathcal{T}$, where $\mathcal{T} = 1.1\,$Hz is a tolerance that is chosen to ensure that all promising configurations are included while maintaining reasonable computation time. Configurations are discarded if they do not satisfy this requirement for one or more of the pairwise couplings. If more than $X_\text{cutoff} = 5000$ possible configurations are identified, only the best $X_\text{cutoff}$ solutions are kept, according to their $\xi$ values.

For the cubic lattice positioning method, the same procedure is followed, with the key difference being that the lattice is adaptively generated depending on the strongest coupling to an already positioned spin in the cluster (see Supplementary Information section V). This ensures that in each case the lattice spacing is fine enough to appropriately sample the volume associated with the dipole-dipole coupling between the nuclear spins.

\noindent\textbf{Data availability}. The data that support the findings of this study are available from the corresponding author upon request.

\noindent\textbf{Acknowledgements.}
We thank V.V. Dobrovitski, W. Hahn, M. Scheer and R. Zia for valuable discussions. We thank R. F. L. Vermeulen and R. N. Schouten for assistance with the RF electronics, and M. Eschen for assistance with the experimental setup. We acknowledge support from the Netherlands Organisation for Scientific Research (NWO) through a Vidi grant.  

\noindent\textbf{Author contributions.} 
MHA and THT devised the experiments. MHA performed the experiments. JR developed the 3D structure analysis. MHA, JR and THT analyzed the data. MHA, JR, MJD and CEB prepared the experimental apparatus. CEB and JR developed the RF electronics. HPB and MAB performed preliminary experiments. MHA, MJD, JR, CEB and THT developed the magnetic field alignment procedure and the $^{14}$N echo spectroscopy. MM and DJT grew the diamond sample. MHA, JR and THT wrote the manuscript with input from all authors. THT supervised the project.


\bibliographystyle{naturemag}
\bibliography{Bib}

\end{document}



\title{Supplementary information for ``Atomic-scale imaging of a 27-nuclear-spin cluster using a single-spin quantum sensor''}

\date{\today}

\maketitle

\section{Properties of the nuclear spin cluster}\label{sec:spinproperties}

This section summarises the properties of the nuclear spin cluster that we retrieved from the 3D spectroscopy. Supplementary Table \ref{table:carbonfrequencies} gives the precession frequencies of the 27 spins that compose the cluster. Supplementary Tables \ref{table:couplingmsm1}-\ref{table:couplingav} summarise the retrieved couplings between these spins. Supplementary Table \ref{table:coordinates} gives the obtained spatial coordinates for the spins in the cluster.

Additionally, Supplementary Table \ref{table:carbonfrequencies} gives estimates for the hyperfine couplings of the nuclear spins to the electron spin. These electron-nuclear couplings play no role in retrieving the structure of the cluster as our imaging method solely relies on the internal nuclear-nuclear couplings. Instead, these couplings are used to estimate realistic bounds on electron-mediated interactions (section \ref{sec:corrections}) and to provide an independent comparison to DFT calculations (section \ref{sec:NV_center_location}).   

The hyperfine couplings are determined as follows. Under the secular approximation and assuming a perfectly aligned magnetic field, the electron-$^{13}$C hyperfine parameters can be calculated as 
\begin{equation}\label{eq:hyperfinefromfreqs}
\begin{split}
A_{\parallel} &= \frac{\omega_{+1}^{2} - \omega_{-1}^{2}}{4\omega_{0}}, \\
A_{\perp} &= \sqrt{\frac{\omega_{+1}^{2}+\omega_{-1}^{2} -2\omega_{0}^{2}-2A_{\parallel}^{2}}{2}}.
\end{split}
\end{equation}
For spins C5, C6, C9, C10, C12, C14, C18, C19, we also measure the precession frequency for $m_s = 0$, while for the rest of the spins we use the average of the measured values ($2\pi \cdot 431.960$ kHz). We observe a standard deviation of 6 Hz in the measured values for $m_s = 0$, which can be attributed to non-secular terms in the Hamiltonian in conjunction with a slightly misaligned magnetic field ($<$ 0.1 deg, see section \ref{sec:bfieldalignment}). For spins C24 to C27, equations \ref{eq:hyperfinefromfreqs} give imaginary values for $A_\perp$, which we attribute to shifts in the $m_s = 0$ precession frequency that are not captured in this approximate model. For these spins we set $A_\perp = 0$ in the table.\\

\begin{table}[h!]
  \centering
  \renewcommand{\arraystretch}{1.5}
  \begin{tabular}{c|c|c|c|c}
  			 & $\omega_{-1} / 2\pi$ (kHz)& $\omega_{+1} / 2\pi$ (kHz) &	$A_\parallel /2\pi$ (kHz) & $A_\perp/2\pi$(kHz) \\
  			  \hline
  			  \hline
   C1	    &	452.83(2)   &   411.40(2)   &	-20.72(1) 	 &	12(1)	\\
   C2	    &	455.37(2)   &   408.956(9)  &	-23.22(1) 	 &	13(1)	\\
   C3       &	463.27(2)   &   400.79(2)   &	-31.25(1) 	 &	8(2)	\\
   C4       &	446.23(4)   &   418.10(1)   &	-14.07(2) 	 &	13(1)	\\ 
   C5	    &   447.234(1)  &   424.752(3)  &   -11.346(2)   &	59.21(3) \\
   C6	    &	480.625(1)  &   383.48(4)   &	-48.58(2) 	 &	9(2)	\\
   C7	    &	440.288(6)  &  423.65(2)    &	 -8.32(1) 	 &	3(5)	\\
   C8	    &	441.77(1)   &  422.20(4)    &	 -9.79(2) 	 &	5(4)	\\
   C9	    &	218.828(1)  & 	 645.123(1) &    213.154(1)  &	3.0(4) 	 \\
   C10	    &	414.407(1)  &    449.687(2) &	 17.643(1) 	 &	8.6(2)	 \\
   C11	    &	417.523(4)  &    446.612(3) &	 14.548(3) 	 &	10(1)	 \\
   C12 	    &	413.477(1)  &    454.427(1) &	 20.569(1) 	 & 	41.51(3) \\
   C13	    &   424.449(1)  &    440.490(1) &	 8.029(1)	 &  21.0(4)	 \\
   C14	    &	 451.802(1) &    412.175(5) &	 -19.815(3)  &	5.3(5)	\\
   C15	    &	 446.010(5) &    418.093(3) &	 -13.961(3)  &	9(1)	\\
   C16	    &	 436.67(5)  &    427.35(3)  &	 -4.66(3) 	 &	7(4)	\\
   C17	    &    437.61(1)  &    426.38(2)  &	 -5.62(1)	 &	5(2)	\\
   C18	    &	469.020(1)  & 	 396.542(1) &   -36.308(1)   &	26.62(4) \\
   C19	    &   408.317(1)  &    457.035(1) &	 24.399(1) 	 &	24.81(4) \\
   C20	    &	429.403(4)   &    434.782(6)&	  2.690(4)	  &	11(1)	 \\
   C21	    &	430.937(3)   &    433.36(1) &	  1.212(5) 	  &	13(1)	\\
   C22       &	424.289(3)   &    439.655(7)&	  7.683(4)    &	4(3)	\\
   C23	    &	435.143(7)   &    428.789(5)&	  -3.177(5)	  &	2(4)	 \\
   C24	    &	436.183(3)   &    427.732(7)&	 -4.225(4)    &	0(6)	\\
   C25	    &	435.827(5)   &    428.079(9)&	  -3.873(5)	  & 0(4)	\\
   C26	    &	435.547(2)   &    428.31(1) &	  -3.618(5)	  &	0(2)	\\
   C27	    &	435.990(3)   &    427.910(9)&	  -4.039(5)	  &	0(3) \\
   \end{tabular}
  \caption{\textbf{The 27 nuclear spins.} Retrieved $^{13}$C spin precession frequencies $\omega_{-1}$, $\omega_{+1}$ for the $m_s = -1$ and $m_s = +1$ electron spin projections respectively. Obtained by least-squares fitting the frequency scan signal in double resonance experiments to a Gaussian function (e.g. Supplementary Fig. \ref{fig:overlapping spins}) or from fits of Ramsey signals to sinusoidal functions with Gaussian decays. $A_\parallel$ and $A_\perp$ are estimates for the parallel and perpendicular components of the electron-$^{13}$C hyperfine interaction respectively, obtained from the measured frequencies $\omega_{-1}$, $\omega_{+1}$ and $\omega_0$ (see equations \ref{eq:hyperfinefromfreqs}). The dephasing time, $T_2^{*}$, for these spins varies from a few milliseconds to 20 ms. The measured coherence time ($T_2$) using a single refocusing pulse is typically $\sim 0.5$ seconds, corresponding to a FWHM spectral resolution of $\sim 1$ Hz in the double resonance experiments.}
  \label{table:carbonfrequencies}
\end{table}

\begin{table}[h!]
\centering
\rotatebox{90}{
\begin{adjustbox}{width=1.2\textwidth}
\renewcommand{\arraystretch}{2.0}
\begin{tabular}{c||cccccccccccccccccccccccccccc}
  &  C1 &  C2 &  C3 &  C4 &  C5 &  C6 &  C7 &  C8 &  C9 &  C10 &  C11 \
&  C12 &  C13 &  C14 &  C15 &  C16 &  C17 &  C18 &  C19 &  C20 &  C21 \
&  C22 &  C23 &  C24 &  C25 &  C26 &  C27 &  N \\
\hline
\hline 
 C1 &  - &  61.90(9) &  61.50(6) &  236.0(2) &  7.05(3) &  12.70(6) \
&   4.8(1) &   2.9(2) &  1.07(7) & $<$1 &  - & $<$1 & $<$1 & $<$1 &  \
- &  - &  - &   1.0(1) & $<$1 &  - &   2.3(2) &  - &  - &   7.8(2) \
&   2.2(2) &  - &  - &  1.95(1) \\
 C2 &  61.90(9) &  - &  236.00(1) &  61.40(8) &  17.20(2) &  19.1(1) \
&  5.61(6) &   1.6(1) &  1.11(6) &   2.1(1) &   1.4(1) &  0.97(9) & \
$<$1 & $<$1 & $<$1 &  - &  - &  1.07(7) & $<$1 &  - &  3.14(7) &  - & \
 - &   4.6(1) &   4.1(2) & $<$1 &  - &  2.25(5) \\
 C3 &  61.50(6) &  236.00(1) &  - &  24.9(1) &  3.21(4) &  29.00(7) \
&  20.2(4) &  13.1(1) &  2.53(7) & $<$1 & $<$1 &  1.37(7) & $<$1 & \
$<$1 &  - &  - &  - &  1.01(7) & $<$1 &  - &   2.3(1) &  - &  - \
&   5.5(1) &   5.0(3) &   3.1(2) &  - &  3.87(9) \\
 C4 &  236.0(2) &  61.40(8) &  24.9(1) &  - &  19.0(1) &   7.0(1) &  \
- &  - &  - &  - &  - & $<$1 & $<$1 & $<$1 & $<$1 &  - &  - &  - &  - \
&  - &  - &  - &  - &  - &  - &  - &  - &  - \\
 C5 &  7.05(3) &  17.20(2) &  3.21(4) &  19.0(1) &  - &   2.5(9) \
&   1.1(2) &   0.9(2) &  0.96(4) &  7.75(3) & $<$1 & $<$1 &  - &  - \
&   1.9(1) &  - &  - &  2.33(5) &   0.6(2) &   7.1(1) &  9.90(7) &  - \
&  - &  - &  - &  - &  - &  - \\
 C6 &  12.70(6) &  19.1(1) &  29.00(7) &   7.0(1) &   2.5(9) &  - \
&  16.50(9) &  12.60(5) &  6.48(2) & $<$1 & $<$1 &  2.01(5) & $<$1 \
&   0.8(2) &  - &  - &  - &  1.28(6) & $<$1 &   1.1(3) &  1.96(9) &  \
- &   0.8(2) &   3.4(1) &   5.6(2) &  4.89(9) &  - &  6.69(4) \\
 C7 &   4.8(1) &  5.61(6) &  20.2(4) &  - &   1.1(2) &  16.50(9) &  - \
&  - &  1.45(7) & $<$1 &  - &   0.4(2) & $<$1 & $<$1 &  - &  - &  - & \
$<$1 &  - &  - &  - &  - &  - &  - &  - &  - &  - &  - \\
 C8 &   2.9(2) &   1.6(1) &  13.1(1) &  - &   0.9(2) &  12.60(5) &  - \
&  - &  1.53(5) & $<$1 &  - &  1.15(9) &  3.11(8) & $<$1 & $<$1 &  - \
&  - & $<$1 &  - &  - &  - &  - &  - &  - &  - &  - &  - &  - \\
 C9 &  1.07(7) &  1.11(6) &  2.53(7) &  - &  0.96(4) &  6.48(2) \
&  1.45(7) &  1.53(5) &  - &  1.12(7) &  1.26(3) &  1.56(4) \
&   1.8(2) &  3.77(3) &  - & $<$1 &   1.1(1) &  1.39(1) &  8.45(2) &  \
- &  - &  3.70(7) &   0.7(1) &  - &  - &   3.5(1) &  0.89(4) \
&  3.25(2) \\
 C10 & $<$1 &   2.1(1) & $<$1 &  - &  7.75(3) & $<$1 & $<$1 & $<$1 \
&  1.12(7) &  - &  14.50(7) &   1.2(1) & $<$1 & $<$1 &   3.7(5) \
&   5.0(1) &  - & $<$1 & $<$1 &   6.9(2) &   3.8(2) &  - &  - &  - &  \
- &  - &  - &  0.56(3) \\
 C11 &  - &   1.4(1) & $<$1 &  - & $<$1 & $<$1 &  - &  - &  1.26(3) \
&  14.50(7) &  - &  3.35(3) & $<$1 & $<$1 &   2.1(3) &   4.7(1) &  - \
&   0.5(3) & $<$1 &   9.5(2) &  - &  - &  - &  - & $<$1 &  - &  - \
&  0.49(2) \\
 C12 & $<$1 &  0.97(9) &  1.37(7) & $<$1 & $<$1 &  2.01(5) &   0.4(2) \
&  1.15(9) &  1.56(4) &   1.2(1) &  3.35(3) &  - &  12.20(6) \
&  1.46(9) &  - & $<$1 &  1.26(8) &   1.1(2) &  0.54(0) &  - &  - \
&   1.7(2) &   1.1(1) &  - &  - &  - &  - &  - \\
 C13 & $<$1 & $<$1 & $<$1 & $<$1 &  - & $<$1 & $<$1 &  3.11(8) \
&   1.8(2) & $<$1 & $<$1 &  12.20(6) &  - & $<$1 & $<$1 & $<$1 & $<$1 \
&  - &   1.3(2) &  - &  - &   7.3(1) &   3.5(1) &  - &  - &  - &  - \
&   1.2(1) \\
 C14 & $<$1 & $<$1 & $<$1 & $<$1 &  - &   0.8(2) & $<$1 & $<$1 \
&  3.77(3) & $<$1 & $<$1 &  1.46(9) & $<$1 &  - &  8.89(7) &   2.1(2) \
&  21.7(3) &  1.39(6) &   1.6(1) &  - &  - &  - &  - &  - &  - &  - \
&   6.3(2) &  0.76(3) \\
 C15 &  - & $<$1 &  - & $<$1 &   1.9(1) &  - &  - & $<$1 &  - \
&   3.7(5) &   2.1(3) &  - & $<$1 &  8.89(7) &  - &  24.6(3) &  - & \
$<$1 & $<$1 &  - &  - &  - &  - &  - &  - &  - &   2.2(6) &  - \\
 C16 &  - &  - &  - &  - &  - &  - &  - &  - & $<$1 &   5.0(1) \
&   4.7(1) & $<$1 & $<$1 &   2.1(2) &  24.6(3) &  - &  - &   1.5(1) & \
$<$1 &  - &  - &  - &  - &  - &  - &  - &  - &  - \\
 C17 &  - &  - &  - &  - &  - &  - &  - &  - &   1.1(1) &  - &  - \
&  1.26(8) & $<$1 &  21.7(3) &  - &  - &  - &  2.42(5) &   0.9(1) &  \
- &  - &  - &  - &  - &  - &  - &  - &  - \\
 C18 &   1.0(1) &  1.07(7) &  1.01(7) &  - &  2.33(5) &  1.28(6) & \
$<$1 & $<$1 &  1.39(1) & $<$1 &   0.5(3) &   1.1(2) &  - &  1.39(6) & \
$<$1 &   1.5(1) &  2.42(5) &  - &  1.32(4) &  - &  1.50(7) &  - &  - \
&  - &  - &  - &  3.60(6) &  1.49(4) \\
 C19 & $<$1 & $<$1 & $<$1 &  - &   0.6(2) & $<$1 &  - &  - &  8.45(2) \
& $<$1 & $<$1 &  0.54(0) &   1.3(2) &   1.6(1) & $<$1 & $<$1 \
&   0.9(1) &  1.32(4) &  - &   1.7(3) &  - &   1.4(1) &  - &  - &  - \
&   0.4(3) &  0.65(5) &  1.72(3) \\
 C20 &  - &  - &  - &  - &   7.1(1) &   1.1(3) &  - &  - &  - \
&   6.9(2) &   9.5(2) &  - &  - &  - &  - &  - &  - &  - &   1.7(3) & \
 - &  - &  - &  - &  - &  - &  - &  - &  - \\
 C21 &   2.3(2) &  3.14(7) &   2.3(1) &  - &  9.90(7) &  1.96(9) &  - \
&  - &  - &   3.8(2) &  - &  - &  - &  - &  - &  - &  - &  1.50(7) &  \
- &  - &  - &  - &  - &  - &  - &  - &  - &  - \\
 C22 &  - &  - &  - &  - &  - &  - &  - &  - &  3.70(7) &  - &  - \
&   1.7(2) &   7.3(1) &  - &  - &  - &  - &  - &   1.4(1) &  - &  - & \
 - &  - &  - &  - &  - &  - &  0.70(7) \\
 C23 &  - &  - &  - &  - &  - &   0.8(2) &  - &  - &   0.7(1) &  - &  \
- &   1.1(1) &   3.5(1) &  - &  - &  - &  - &  - &  - &  - &  - &  - \
&  - &  - &  - &  - &  - &  - \\
 C24 &   7.8(2) &   4.6(1) &   5.5(1) &  - &  - &   3.4(1) &  - &  - \
&  - &  - &  - &  - &  - &  - &  - &  - &  - &  - &  - &  - &  - &  - \
&  - &  - &  - &  - &  - &  - \\
 C25 &   2.2(2) &   4.1(2) &   5.0(3) &  - &  - &   5.6(2) &  - &  - \
&  - &  - & $<$1 &  - &  - &  - &  - &  - &  - &  - &  - &  - &  - &  \
- &  - &  - &  - &  - &  - &  - \\
 C26 &  - & $<$1 &   3.1(2) &  - &  - &  4.89(9) &  - &  - &   3.5(1) \
&  - &  - &  - &  - &  - &  - &  - &  - &  - &   0.4(3) &  - &  - &  \
- &  - &  - &  - &  - &  - &  - \\
 C27 &  - &  - &  - &  - &  - &  - &  - &  - &  0.89(4) &  - &  - &  \
- &  - &   6.3(2) &   2.2(6) &  - &  - &  3.60(6) &  0.65(5) &  - &  \
- &  - &  - &  - &  - &  - &  - &  - \\
 N &  1.95(1) &  2.25(5) &  3.87(9) &  - &  - &  6.69(4) &  - &  - \
&  3.25(2) &  0.56(3) &  0.49(2) &  - &   1.2(1) &  0.76(3) &  - &  - \
&  - &  1.49(4) &  1.72(3) &  - &  - &  0.70(7) &  - &  - &  - &  - & \
 - &  - \\
\end{tabular} 
\end{adjustbox}
}
\caption{All measured coupling frequencies in Hz for the $m_s = -1$ electron spin projection. To account for the cases where pulse errors cannot be neglected, the coupling frequencies are extracted by fitting the time-domain double resonance signals to $S  = a + A\cdot e^{-(t/T_2)^n}\cos{(2\pi{f} t+ \phi)} + B\cdot e^{-(t/T_2)^n}$, where $T_2$ is the coherence time and $n$, $A$ and $B$ are fit parameters that account for the signal decay shape, contrast and pulse errors. For the couplings marked as $< 1\,$Hz in the tables, no oscillation was observed within the decay time (for these measurements, $N=1$ and $T_2 \sim 0.5$ seconds).}
\label{table:couplingmsm1}
\end{table}

\clearpage

\begin{table}[h!]
\centering
\rotatebox{90}{
\begin{adjustbox}{width=1.2\textwidth}
\renewcommand{\arraystretch}{2.0}
\begin{tabular}{c||cccccccccccccccccccccccccccc}
  &  C1 &  C2 &  C3 &  C4 &  C5 &  C6 &  C7 &  C8 &  C9 &  C10 &  C11 \
&  C12 &  C13 &  C14 &  C15 &  C16 &  C17 &  C18 &  C19 &  C20 &  C21 \
&  C22 &  C23 &  C24 &  C25 &  C26 &  C27 &  N \\
\hline
\hline
 C1 &  - &  61.90(7) &  62.90(9) &  236.0(2) &  10.1(1) &  12.70(7) \
&   5.9(2) &   2.6(2) &   1.1(2) &  - &  - &  - &  - &  - &  - &  - & \
 - &   1.1(3) &  - &  - &   2.7(5) &  - &  - &   7.5(3) &   1.8(3) &  \
- &  - &  1.86(3) \\
 C2 &  61.90(7) &  - &  236.0(1) &  62.7(2) &  22.7(2) &  19.30(3) \
&   6.3(3) &   2.2(3) &  1.00(9) &  - &  1.40(0) & $<$1 &  - &  - &  \
- &  - &  - &   1.0(2) &  - &  - &  2.60(0) &  - &  - &   4.8(1) \
&  3.35(9) &   1.2(6) &  - &  2.06(4) \\
 C3 &  62.90(9) &  236.0(1) &  - &  25.2(1) &   7.1(2) &  24.8(1) \
&  20.10(6) &  12.30(5) &  2.46(5) &  - &  - &  1.14(8) &  - &  - &  \
- &  - &  - &  0.97(0) &  - &  - &   1.8(3) &  - &  - &  5.43(6) \
&   4.6(1) &   3.1(2) &  - &  - \\
 C4 &  236.0(2) &  62.7(2) &  25.2(1) &  - &  20.2(4) &   7.2(2) &  - \
&  - &  - &  - &  - &  - &  - &  - &  - &  - &  - &  - &  - &  - &  - \
&  - &  - &  - &  - &  - &  - &  - \\
 C5 &  10.1(1) &  22.7(2) &   7.1(2) &  20.2(4) &  - &   1.1(3) \
&   1.4(8) &   1.7(1) &  1.35(9) &   9.9(1) &  - &  - &  1.86(9) &  - \
&  - &  - &  - &   1.3(2) &  1.05(8) &   7.9(2) &  15.3(1) &  - &  - \
&  - &  - &  - &  - &  - \\
 C6 &  12.70(7) &  19.30(3) &  24.8(1) &   7.2(2) &   1.1(3) &  - \
&  16.70(4) &  12.1(1) &  6.74(4) &  - &  - &  1.74(6) &  - \
&  1.04(7) &  - &  - &  - &  1.07(6) &   0.7(1) & $<$1 &   1.6(2) &  \
- &  0.78(8) &  3.36(6) &  3.37(7) &  5.09(6) &  - &  6.60(7) \\
 C7 &   5.9(2) &   6.3(3) &  20.10(6) &  - &   1.4(8) &  16.70(4) &  \
- &  - &  1.41(6) &  - &  - &   0.4(2) &  - &  - &  - &  - &  - &  - \
&  - &  - &  - &  - &  - &  - &  - &  - &  - &  - \\
 C8 &   2.6(2) &   2.2(3) &  12.30(5) &  - &   1.7(1) &  12.1(1) &  - \
&  - &  1.65(5) &  - &  - &  1.10(7) &  3.61(6) &  - &  - &  - &  - & \
 - &  - &  - &  - &  - &  - &  - &  - &  - &  - &  - \\
 C9 &   1.1(2) &  1.00(9) &  2.46(5) &  - &  1.35(9) &  6.74(4) \
&  1.41(6) &  1.65(5) &  - &   1.2(2) &  1.23(6) &  3.48(7) \
&  1.96(7) &  3.80(6) &  - &  - &   0.9(1) &  1.16(6) &  9.57(2) &  - \
&  - &  3.82(6) &   0.8(2) &  - &  - &   3.1(1) &  0.94(4) &  3.24(3) \
\\
 C10 &  - &  - &  - &  - &   9.9(1) &  - &  - &  - &   1.2(2) &  - \
&  14.50(6) &   1.2(1) &  - &  - &   3.7(1) &  - &  - &  - &   0.7(2) \
&   6.9(3) &  - &  - &  - &  - &  - &  - &  - &  - \\
 C11 &  - &  1.40(0) &  - &  - &  - &  - &  - &  - &  1.23(6) \
&  14.50(6) &  - &  3.38(6) &  - &  - &   1.7(5) &  - &  - &   0.5(3) \
&  - &  - &  - &  - &  - &  - &  - &  - &  - &  - \\
 C12 &  - & $<$1 &  1.14(8) &  - &  - &  1.74(6) &   0.4(2) \
&  1.10(7) &  3.48(7) &   1.2(1) &  3.38(6) &  - &  12.20(3) \
&  2.48(9) &  - &  - &  1.42(9) &   1.1(5) &  0.81(7) &  - &  - \
&  2.35(4) &   1.1(2) &  - &  - &  - &  - &  - \\
 C13 &  - &  - &  - &  - &  1.86(9) &  - &  - &  3.61(6) &  1.96(7) & \
 - &  - &  12.20(3) &  - &  - &  - &  - &  - &  - &  0.83(8) &  - &  \
- &  7.21(4) &   3.5(1) &  - &   1.5(1) &  - &  - &  1.06(2) \\
 C14 &  - &  - &  - &  - &  - &  1.04(7) &  - &  - &  3.80(6) &  - &  \
- &  2.48(9) &  - &  - &  8.86(1) &  2.13(9) &  22.00(6) &   1.0(2) \
&  1.61(7) &  - &  - &  - &  - &  - &  - &  - &   6.6(1) &  - \\
 C15 &  - &  - &  - &  - &  - &  - &  - &  - &  - &   3.7(1) \
&   1.7(5) &  - &  - &  8.86(1) &  - &  26.2(5) &   1.7(2) &  - &  - \
&  - &  - &  - &  - &  - &  - &  - &   2.7(2) &  - \\
 C16 &  - &  - &  - &  - &  - &  - &  - &  - &  - &  - &  - &  - &  - \
&  2.13(9) &  26.2(5) &  - &  - &  1.39(7) &  - &  - &  - &  - &  - & \
 - &  - &  - &  - &  - \\
 C17 &  - &  - &  - &  - &  - &  - &  - &  - &   0.9(1) &  - &  - \
&  1.42(9) &  - &  22.00(6) &   1.7(2) &  - &  - &  2.70(8) \
&   1.0(1) &  - &  - &  - &  - &  - &  - &  - &  - &  - \\
 C18 &   1.1(3) &   1.0(2) &  0.97(0) &  - &   1.3(2) &  1.07(6) &  - \
&  - &  1.16(6) &  - &   0.5(3) &   1.1(5) &  - &   1.0(2) &  - \
&  1.39(7) &  2.70(8) &  - &  0.95(6) &  - &   1.3(1) &  - &  - &  - \
&  - &  - &  3.56(9) &  - \\
 C19 &  - &  - &  - &  - &  1.05(8) &   0.7(1) &  - &  - &  9.57(2) \
&   0.7(2) &  - &  0.81(7) &  0.83(8) &  1.61(7) &  - &  - &   1.0(1) \
&  0.95(6) &  - & $<$1 &  - &   1.3(1) &  - &  - &  - & $<$1 \
&  0.88(4) &  - \\
 C20 &  - &  - &  - &  - &   7.9(2) & $<$1 &  - &  - &  - &   6.9(3) \
&  - &  - &  - &  - &  - &  - &  - &  - & $<$1 &  - &  - &  - &  - &  \
- &  - &  - &  - &  - \\
 C21 &   2.7(5) &  2.60(0) &   1.8(3) &  - &  15.3(1) &   1.6(2) &  - \
&  - &  - &  - &  - &  - &  - &  - &  - &  - &  - &   1.3(1) &  - &  \
- &  - &  - &  - &  - &  - &  - &  - &  - \\
 C22 &  - &  - &  - &  - &  - &  - &  - &  - &  3.82(6) &  - &  - \
&  2.35(4) &  7.21(4) &  - &  - &  - &  - &  - &   1.3(1) &  - &  - & \
 - &  - &  - &  - &  - &  - &  - \\
 C23 &  - &  - &  - &  - &  - &  0.78(8) &  - &  - &   0.8(2) &  - &  \
- &   1.1(2) &   3.5(1) &  - &  - &  - &  - &  - &  - &  - &  - &  - \
&  - &  - &  - &  - &  - &  - \\
 C24 &   7.5(3) &   4.8(1) &  5.43(6) &  - &  - &  3.36(6) &  - &  - \
&  - &  - &  - &  - &  - &  - &  - &  - &  - &  - &  - &  - &  - &  - \
&  - &  - &  - &  - &  - &  - \\
 C25 &   1.8(3) &  3.35(9) &   4.6(1) &  - &  - &  3.37(7) &  - &  - \
&  - &  - &  - &  - &   1.5(1) &  - &  - &  - &  - &  - &  - &  - &  \
- &  - &  - &  - &  - &  - &  - &  - \\
 C26 &  - &   1.2(6) &   3.1(2) &  - &  - &  5.09(6) &  - &  - \
&   3.1(1) &  - &  - &  - &  - &  - &  - &  - &  - &  - & $<$1 &  - & \
 - &  - &  - &  - &  - &  - &  - &  - \\
 C27 &  - &  - &  - &  - &  - &  - &  - &  - &  0.94(4) &  - &  - &  \
- &  - &   6.6(1) &   2.7(2) &  - &  - &  3.56(9) &  0.88(4) &  - &  \
- &  - &  - &  - &  - &  - &  - &  - \\
 N &  1.86(3) &  2.06(4) &  - &  - &  - &  6.60(7) &  - &  - \
&  3.24(3) &  - &  - &  - &  1.06(2) &  - &  - &  - &  - &  - &  - &  \
- &  - &  - &  - &  - &  - &  - &  - &  - \\
\end{tabular}
\end{adjustbox}
}
\caption{All measured coupling frequencies in Hz for the $m_s = +1$ electron spin projection.}
\label{table:couplingmsp1}
\end{table}
\clearpage

\begin{table}[h!]
\centering
\rotatebox{90}{
\begin{adjustbox}{width=1.2\textwidth}
\renewcommand{\arraystretch}{2.0}
\begin{tabular}{c||cccccccccccccccccccccccccccc}
  &  C1 &  C2 &  C3 &  C4 &  C5 &  C6 &  C7 &  C8 &  C9 &  C10 &  C11 \
&  C12 &  C13 &  C14 &  C15 &  C16 &  C17 &  C18 &  C19 &  C20 &  C21 \
&  C22 &  C23 &  C24 &  C25 &  C26 &  C27 &  N \\
\hline
\hline
 C1 &  - &  61.90(8) &  62.20(8) &  236.0(2) &  8.57(7) &  12.70(7) \
&   5.4(2) &   2.8(2) &   1.1(2) & $<$1 &  - & $<$1 & $<$1 & $<$1 &  \
- &  - &  - &   1.1(2) & $<$1 &  - &   2.5(4) &  - &  - &   7.7(2) \
&   2.0(2) &  - &  - &  1.90(2) \\
 C2 &  61.90(8) &  - &  236.00(7) &  62.1(2) &  19.9(1) &  19.20(7) \
&   5.9(2) &   1.9(2) &  1.06(8) &   2.1(1) &  1.40(7) &  0.97(9) & \
$<$1 & $<$1 & $<$1 &  - &  - &   1.0(2) & $<$1 &  - &  2.87(5) &  - & \
 - &   4.7(1) &   3.7(2) &   1.2(6) &  - &  2.16(5) \\
 C3 &  62.20(8) &  236.00(7) &  - &  25.1(1) &   5.2(1) &  26.90(9) \
&  20.1(3) &  12.70(8) &  2.50(6) & $<$1 & $<$1 &  1.25(8) & $<$1 & \
$<$1 &  - &  - &  - &  0.99(5) & $<$1 &  - &   2.0(2) &  - &  - \
&  5.46(8) &   4.8(2) &   3.1(2) &  - &  3.87(9) \\
 C4 &  236.0(2) &  62.1(2) &  25.1(1) &  - &  19.6(3) &   7.1(2) &  - \
&  - &  - &  - &  - & $<$1 & $<$1 & $<$1 & $<$1 &  - &  - &  - &  - & \
 - &  - &  - &  - &  - &  - &  - &  - &  - \\
 C5 &  8.57(7) &  19.9(1) &   5.2(1) &  19.6(3) &  - &   1.8(7) \
&   1.3(6) &   1.3(2) &  1.16(7) &  8.82(7) & $<$1 & $<$1 &  1.86(9) \
&  - &   1.9(1) &  - &  - &   1.8(2) &   0.8(2) &   7.5(2) \
&  12.60(9) &  - &  - &  - &  - &  - &  - &  - \\
 C6 &  12.70(7) &  19.20(7) &  26.90(9) &   7.1(2) &   1.8(7) &  - \
&  16.60(7) &  12.40(8) &  6.61(3) & $<$1 & $<$1 &  1.88(6) & $<$1 \
&   0.9(2) &  - &  - &  - &  1.18(6) &   0.7(1) &   1.1(3) &   1.8(2) \
&  - &   0.8(2) &  3.38(8) &   4.5(2) &  4.99(8) &  - &  6.64(6) \\
 C7 &   5.4(2) &   5.9(2) &  20.1(3) &  - &   1.3(6) &  16.60(7) &  - \
&  - &  1.43(7) & $<$1 &  - &   0.4(2) & $<$1 & $<$1 &  - &  - &  - & \
$<$1 &  - &  - &  - &  - &  - &  - &  - &  - &  - &  - \\
 C8 &   2.8(2) &   1.9(2) &  12.70(8) &  - &   1.3(2) &  12.40(8) &  \
- &  - &  1.59(5) & $<$1 &  - &  1.12(8) &  3.36(7) & $<$1 & $<$1 &  \
- &  - & $<$1 &  - &  - &  - &  - &  - &  - &  - &  - &  - &  - \\
 C9 &   1.1(2) &  1.06(8) &  2.50(6) &  - &  1.16(7) &  6.61(3) \
&  1.43(7) &  1.59(5) &  - &   1.2(2) &  1.25(5) &  2.52(6) \
&   1.9(2) &  3.78(5) &  - & $<$1 &   1.0(1) &  1.27(4) &  9.01(2) &  \
- &  - &  3.76(7) &   0.8(2) &  - &  - &   3.3(1) &  0.92(4) \
&  3.24(3) \\
 C10 & $<$1 &   2.1(1) & $<$1 &  - &  8.82(7) & $<$1 & $<$1 & $<$1 \
&   1.2(2) &  - &  14.50(7) &   1.2(1) & $<$1 & $<$1 &   3.7(4) \
&   5.0(1) &  - & $<$1 &   0.7(2) &   6.9(2) &   3.8(2) &  - &  - &  \
- &  - &  - &  - &  0.56(3) \\
 C11 &  - &  1.40(7) & $<$1 &  - & $<$1 & $<$1 &  - &  - &  1.25(5) \
&  14.50(7) &  - &  3.36(5) & $<$1 & $<$1 &   1.9(4) &   4.7(1) &  - \
&   0.5(3) & $<$1 &   9.5(2) &  - &  - &  - &  - & $<$1 &  - &  - \
&  0.49(2) \\
 C12 & $<$1 &  0.97(9) &  1.25(8) & $<$1 & $<$1 &  1.88(6) &   0.4(2) \
&  1.12(8) &  2.52(6) &   1.2(1) &  3.36(5) &  - &  12.20(5) \
&  1.97(9) &  - & $<$1 &  1.34(9) &   1.1(4) &  0.67(5) &  - &  - \
&   2.0(1) &   1.1(2) &  - &  - &  - &  - &  - \\
 C13 & $<$1 & $<$1 & $<$1 & $<$1 &  1.86(9) & $<$1 & $<$1 &  3.36(7) \
&   1.9(2) & $<$1 & $<$1 &  12.20(5) &  - & $<$1 & $<$1 & $<$1 & $<$1 \
&  - &   1.1(2) &  - &  - &  7.26(8) &   3.5(1) &  - &   1.5(1) &  - \
&  - &  1.13(7) \\
 C14 & $<$1 & $<$1 & $<$1 & $<$1 &  - &   0.9(2) & $<$1 & $<$1 \
&  3.78(5) & $<$1 & $<$1 &  1.97(9) & $<$1 &  - &  8.88(5) &   2.1(2) \
&  21.9(2) &   1.2(2) &  1.60(9) &  - &  - &  - &  - &  - &  - &  - \
&   6.5(2) &  0.76(3) \\
 C15 &  - & $<$1 &  - & $<$1 &   1.9(1) &  - &  - & $<$1 &  - \
&   3.7(4) &   1.9(4) &  - & $<$1 &  8.88(5) &  - &  25.4(4) \
&   1.7(2) & $<$1 & $<$1 &  - &  - &  - &  - &  - &  - &  - \
&   2.4(4) &  - \\
 C16 &  - &  - &  - &  - &  - &  - &  - &  - & $<$1 &   5.0(1) \
&   4.7(1) & $<$1 & $<$1 &   2.1(2) &  25.4(4) &  - &  - &  1.44(9) & \
$<$1 &  - &  - &  - &  - &  - &  - &  - &  - &  - \\
 C17 &  - &  - &  - &  - &  - &  - &  - &  - &   1.0(1) &  - &  - \
&  1.34(9) & $<$1 &  21.9(2) &   1.7(2) &  - &  - &  2.56(7) \
&   1.0(1) &  - &  - &  - &  - &  - &  - &  - &  - &  - \\
 C18 &   1.1(2) &   1.0(2) &  0.99(5) &  - &   1.8(2) &  1.18(6) & \
$<$1 & $<$1 &  1.27(4) & $<$1 &   0.5(3) &   1.1(4) &  - &   1.2(2) & \
$<$1 &  1.44(9) &  2.56(7) &  - &  1.14(5) &  - &  1.40(9) &  - &  - \
&  - &  - &  - &  3.58(8) &  1.49(4) \\
 C19 & $<$1 & $<$1 & $<$1 &  - &   0.8(2) &   0.7(1) &  - &  - \
&  9.01(2) &   0.7(2) & $<$1 &  0.67(5) &   1.1(2) &  1.60(9) & $<$1 \
& $<$1 &   1.0(1) &  1.14(5) &  - &   1.7(3) &  - &   1.4(1) &  - &  \
- &  - &   0.4(3) &  0.76(5) &  1.72(3) \\
 C20 &  - &  - &  - &  - &   7.5(2) &   1.1(3) &  - &  - &  - \
&   6.9(2) &   9.5(2) &  - &  - &  - &  - &  - &  - &  - &   1.7(3) & \
 - &  - &  - &  - &  - &  - &  - &  - &  - \\
 C21 &   2.5(4) &  2.87(5) &   2.0(2) &  - &  12.60(9) &   1.8(2) &  \
- &  - &  - &   3.8(2) &  - &  - &  - &  - &  - &  - &  - &  1.40(9) \
&  - &  - &  - &  - &  - &  - &  - &  - &  - &  - \\
 C22 &  - &  - &  - &  - &  - &  - &  - &  - &  3.76(7) &  - &  - \
&   2.0(1) &  7.26(8) &  - &  - &  - &  - &  - &   1.4(1) &  - &  - & \
 - &  - &  - &  - &  - &  - &  0.70(7) \\
 C23 &  - &  - &  - &  - &  - &   0.8(2) &  - &  - &   0.8(2) &  - &  \
- &   1.1(2) &   3.5(1) &  - &  - &  - &  - &  - &  - &  - &  - &  - \
&  - &  - &  - &  - &  - &  - \\
 C24 &   7.7(2) &   4.7(1) &  5.46(8) &  - &  - &  3.38(8) &  - &  - \
&  - &  - &  - &  - &  - &  - &  - &  - &  - &  - &  - &  - &  - &  - \
&  - &  - &  - &  - &  - &  - \\
 C25 &   2.0(2) &   3.7(2) &   4.8(2) &  - &  - &   4.5(2) &  - &  - \
&  - &  - & $<$1 &  - &   1.5(1) &  - &  - &  - &  - &  - &  - &  - & \
 - &  - &  - &  - &  - &  - &  - &  - \\
 C26 &  - &   1.2(6) &   3.1(2) &  - &  - &  4.99(8) &  - &  - \
&   3.3(1) &  - &  - &  - &  - &  - &  - &  - &  - &  - &   0.4(3) &  \
- &  - &  - &  - &  - &  - &  - &  - &  - \\
 C27 &  - &  - &  - &  - &  - &  - &  - &  - &  0.92(4) &  - &  - &  \
- &  - &   6.5(2) &   2.4(4) &  - &  - &  3.58(8) &  0.76(5) &  - &  \
- &  - &  - &  - &  - &  - &  - &  - \\
 N &  1.90(2) &  2.16(5) &  3.87(9) &  - &  - &  6.64(6) &  - &  - \
&  3.24(3) &  0.56(3) &  0.49(2) &  - &  1.13(7) &  0.76(3) &  - &  - \
&  - &  1.49(4) &  1.72(3) &  - &  - &  0.70(7) &  - &  - &  - &  - & \
 - &  - \\
\end{tabular}
\end{adjustbox}
}
\caption{All measured coupling frequencies in Hz averaged over the $m_s = \pm1$ electron spin projections. A total of 171 couplings are measured, including the couplings marked $< 1\,$Hz.}
\label{table:couplingav}
\end{table}
\clearpage

\begin{table}[h!]
\centering
\renewcommand{\arraystretch}{1.5}
\begin{tabular}{c|ccc|ccc|ccc|ccc}
\text{Spin } & \multicolumn{3}{c|}{Diamond} & \multicolumn{3}{c|}{Diamond Fit} & \multicolumn{3}{c|}{Cubic} & \multicolumn{3}{c}{Cubic Fit}  \\ \hline \hline
  & x (\AA) & y (\AA) & z (\AA) & x (\AA) & y (\AA) & z (\AA) & x (\AA) & y (\AA) & z (\AA) & x (\AA) & y (\AA) & z (\AA) \\ \hline
 1 &   0.00 &   0.00 &   0.00 &  0.00$^\dagger$ &  0.00$^\dagger$ &  0.00$^\dagger$ \
&   0.00 &   0.00 &   0.00 &  0.00$^\dagger$ &  0.00$^\dagger$ &  0.00$^\dagger$ \\
 2 &   2.52 &   2.91 &  -0.51 &  2.53(2)$^\ddagger$ &  2.92(2)$^\ddagger$& -0.45(7) \
&   2.52 &   2.91 &  -0.50 &  2.52(2)$^\ddagger$ &  2.91(2)$^\ddagger$ & -0.47(7) \\
 3 &   3.78 &   0.73 &  -0.51 &  3.77(3) &  0.72(3) & -0.50(9) \
&   3.75 &   0.71 &  -0.55 &  3.78(3) &  0.72(3) & -0.48(9) \\
 4 &  -1.26 &   2.18 &   0.00 & -1.28(3) &  2.18(3) & - 0.0(1) &  \
-1.23 &   2.20 &   0.05 & -1.28(3) &  2.17(4) & - 0.1(1) \\
 5 &   0.00 &   4.37 &  -6.18 &  0.06(5) &  4.45(5) & -6.17(7) \
&   0.03 &   4.38 &  -6.19 &  0.05(5) &  4.45(5) & -6.18(7) \\
 6 &   5.04 &  -1.46 &  -2.06 &   5.1(1) & - 1.4(1) & -2.03(6) \
&   5.14 &  -1.36 &  -2.08 &   5.1(1) & - 1.4(1) & -2.02(6) \\
 7 &   5.04 &  -1.46 &   5.66 &  4.93(9) & - 1.5(1) &  5.67(8) \
&   4.83 &  -1.59 &   5.65 &  4.92(9) & - 1.6(1) &  5.67(8) \\
 8 &   7.57 &   1.46 &   3.60 &   7.5(2) &   1.7(2) &   3.6(1) \
&   7.42 &   1.72 &   3.32 &   7.5(2) &   1.6(2) &   3.6(1) \\
 9 &   7.57 &  -4.37 & -10.81 &   7.3(4) & - 4.6(5) & -10.7(2) \
&   7.70 &  -4.31 & -10.90 &   7.2(4) & - 4.5(4) & -11.0(2) \\
 10 &   0.00 &   8.74 & -12.36 &   0.0(3) &   8.5(3) & -13.2(3) \
&   0.05 &   8.56 & -12.99 & - 0.1(3) &   8.6(3) & -13.1(3) \\
 11 &   6.31 &   9.46 & -12.87 &   6.3(5) &   9.5(4) & -12.6(3) \
&   6.39 &   9.47 & -12.67 &   6.2(4) &   9.6(4) & -12.7(3) \\
 12 &  11.35 &   0.73 & -14.42 &  11.4(5) &   0.9(5) & -14.5(2) \
&  11.37 &   0.54 & -14.42 &  11.0(5) &   1.0(5) & -14.7(2) \\
 13 &  12.61 &   2.91 &  -6.69 &  12.7(6) &   3.3(6) & - 6.9(2) \
&  12.23 &   3.23 &  -6.99 &  12.3(6) &   3.3(5) & - 7.1(2) \\
 14 &   5.04 &  -2.91 & -22.65 &   5.4(4) & - 3.2(4) & -22.9(3) \
&   5.40 &  -3.35 & -22.72 &   4.9(4) & - 3.2(4) & -23.2(3) \\
 15 &   1.26 &   3.64 & -22.65 &   2.0(5) &   3.5(4) & -22.7(4) \
&   2.34 &   3.48 & -22.27 &   1.8(5) &   3.6(4) & -22.7(4) \\
 16 &   2.52 &   8.74 & -23.17 &   3.5(6) &   8.4(5) & -23.5(3) \
&   3.26 &   8.28 & -23.38 &   3.3(5) &   8.5(5) & -23.5(3) \\
 17 &   6.31 &  -2.18 & -29.34 &   5.8(8) & - 2.6(8) & -29.8(4) \
&   5.40 &  -3.35 & -29.75 &   5.7(7) & - 2.7(8) & -30.1(4) \\
 18 &   0.00 &  -1.46 & -19.05 & - 0.2(4) & - 1.2(5) & -19.1(3) \
&   0.17 &  -1.55 & -18.46 & - 0.2(4) & - 1.3(4) & -19.0(3) \\
 19 &   3.78 &  -9.46 &  -8.75 &   3.8(8) & - 9.1(8) & - 8.4(4) \
&   4.55 &  -9.32 &  -8.65 &   3.3(8) & - 9.4(8) & - 8.8(5) \\
 20 &   3.78 &  10.92 &  -4.63 &   3.1(5) &  11.8(5) & - 5.7(6) \
&   3.24 &  11.64 &  -5.49 &   3.0(5) &  11.9(4) & - 5.7(6) \\
 21 &  -5.04 &   5.82 &  -4.12 & - 4.7(4) &   5.8(5) & -4.02(8) &  \
-4.81 &   5.78 &  -4.08 & - 4.8(4) &   5.7(4) & -4.03(8) \\
 22 &  16.39 &  -3.64 &  -8.24 &  16(1) & -3(1) & -8(1) &  16.94 &  \
-2.19 &  -9.02 &  15(1) & -3(1) & - 8.4(7) \\
 23 &  16.39 &  -0.73 &   3.60 &  16(1) &  0(1) &   3.5(6) &  16.14 & \
 -0.16 &   3.37 &  15(1) &  0(1) &   3.5(5) \\
 24 &   1.26 &  -0.73 &   9.78 &   1.2(3) & - 0.8(3) &   9.8(1) \
&   1.08 &  -0.28 &   9.98 &   1.2(3) & - 0.8(3) &   9.8(1) \\
 25 &   6.31 &  -0.73 &   9.78 &   7.5(4) &   0.5(5) &   9.4(3) \
&   7.17 &   0.47 &   9.74 &   7.5(4) &   0.5(5) &   9.4(3) \\
 26 &  12.61 &  -5.82 &  -0.51 &  12.6(8) & - 6.1(9) & - 0.7(6) \
&  13.12 &  -5.59 &  -1.40 &  12.3(8) & - 6.6(9) & - 0.8(6) \\
 27 &   1.26 &  -3.64 & -31.40 &   1.3(4) & - 3.3(4) & -31.4(4) \
&   1.06 &  -3.27 & -31.06 &   0.7(4) & - 3.1(4) & -31.5(4) \\
 N &   3.78 &  -0.73 &  -8.75 &   3.8(3)* & - 0.7(3)* & -8.61(8)* & - & \
- & - & - & - & - \\
\end{tabular}
\caption{\textbf{Structure of the cluster.} Coordinates obtained from the measured couplings using the diamond lattice positioning method (section \ref{sec:diamondlattice}), the cubic lattice method (section \ref{sec:cubiclattice}) and from using least-squares minimisation using the diamond and cubic solutions as an initial guess (section \ref{sec:fit}). $^\dagger$Coordinate fixed to zero in the fitting routine. $^\ddagger$Rotational symmetry fixed by rotating the initial guess solution by $-49.1\,$deg and fixing the rotated coordinate $y_2'$ to zero. *Nitrogen spin fitted coordinates when the $^{13}$C coordinates are fixed to the diamond lattice solution (see section \ref{sec:NV_center_location}). }
\label{table:coordinates}
\end{table}
\clearpage

\begin{figure}[h!]
\centering

\includegraphics[width=1.0\textwidth]{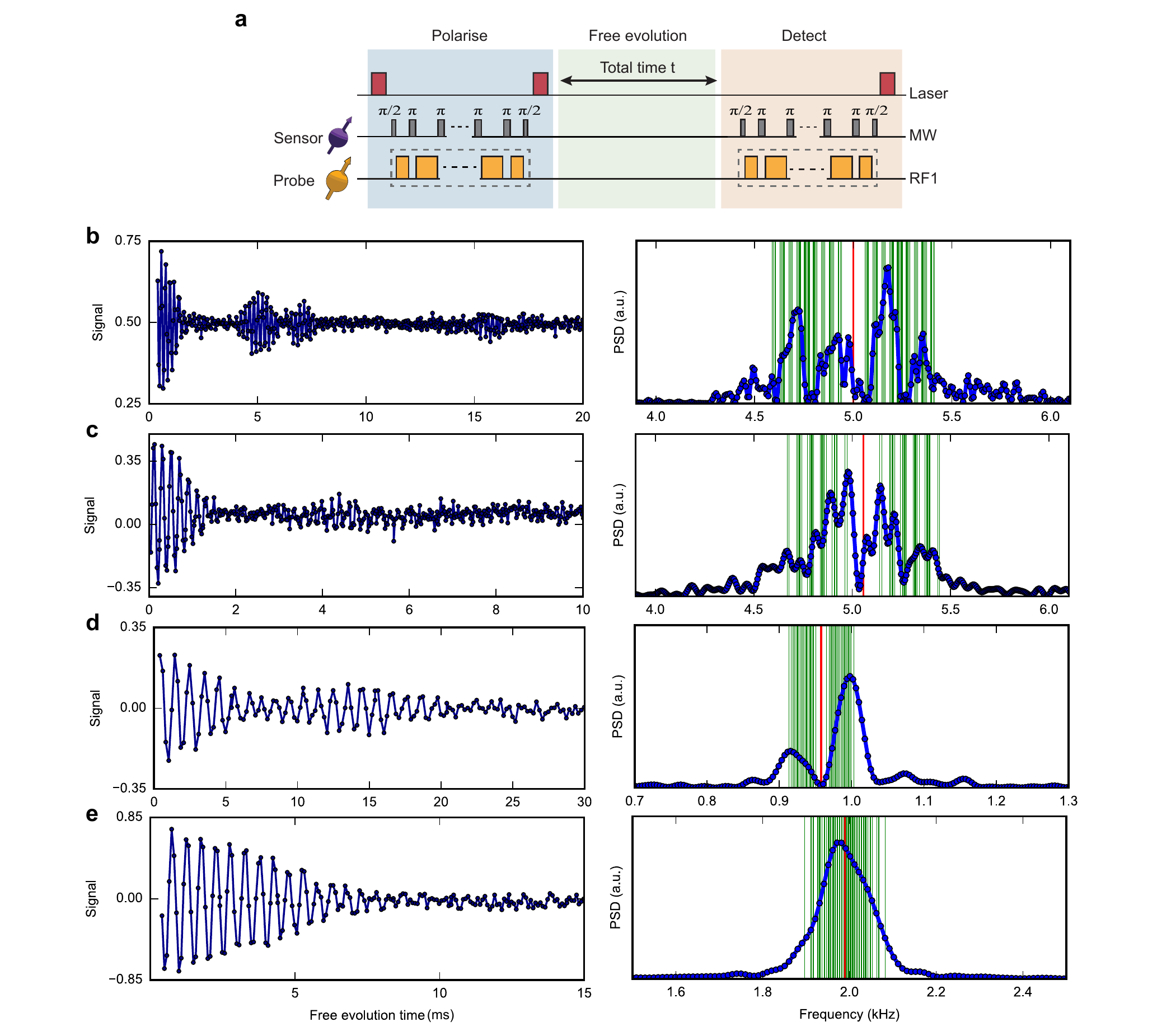}

\caption{{\bf Ramsey experiments and underlying spectra for different nuclear spins in the cluster.} \textbf{a}, Schematic of the pulse sequence used to perform the Ramsey experiment (see main text). \textbf{b}, Ramsey signal as a function of free evolution time for C2 (measured with a 5 kHz detuning), with the corresponding power spectral density of the obtained signal, revealing a complex spectrum. This gives a first indication of the presence of nuclear-nuclear spin interactions. The red line represents the central frequency $f_0$. Green lines are the $2^7$ frequencies based on the 7 strongest coupling strengths extracted from our high resolution double resonance spectroscopy (Supplementary Table \ref{table:couplingav}). These frequencies are given by $f_0 \pm f_1 \pm f_2 \pm f_3 \pm f_4 \pm f_5 \pm f_6 \pm f_7 $, where $f_{1}$ to $f_{7}$ are the 7 largest measured coupling strengths for C2. This comparison highlights the large effective improvement in resolution, and the ability to resolve dense spectra, of our method. \textbf{c}, The same experiment for C3 ($\sim$ 5 kHz detuning), \textbf{d}, for C15 ($\sim$ 1 kHz detuning) and \textbf{e}, for C5 ($\sim$ 2 kHz detuning).  }
\label{fig:Ramseys}
\end{figure}

\clearpage

\section{ Multi-resonance experiments and resolving spectrally overlapping spins}\label{sec:multiresonance}

As discussed in the main text, resolving ambiguities due to overlapping signals from multiple spins at (near-)identical frequencies is a key component for determining the composition of the cluster and how the spins couple to each other. To resolve such ambiguities, there are two main challenges: first, how to extract the underlying coupling frequencies from the complex signals in the case of multiple overlapping spins; second, how to determine the number of spins in the cluster and to assign the measured couplings to specific spins.

\begin{figure}[h!]
\centering

\includegraphics[width=0.8\textwidth]{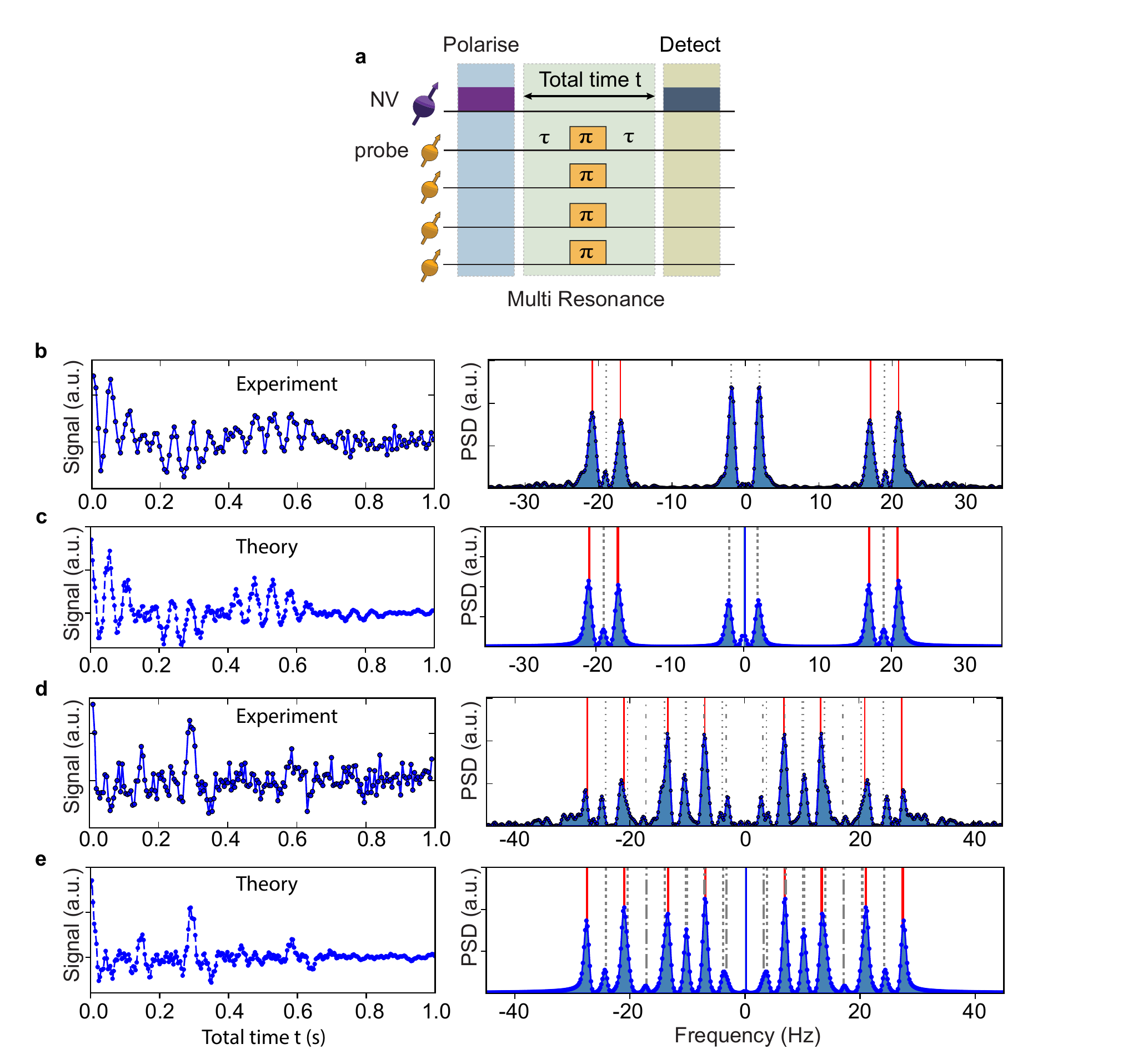}

\caption{{\bf Multi-resonance experiments: probing multiple couplings simultaneously.} \textbf{a}, A pulse sequence describing a multi-resonance experiment that causes multiple couplings to be accessed simultaneously. The echos are performed simultaneously on the probe and the $N$ target spins. This can be realised using $N$ resonant RF pulses or a single spectrally broad pulse. When the NMR lines of multiple spins overlap, the last situation naturally and inevitably occurs . \textbf{b}, In this case two spectrally overlapping spins in the cluster are flipped simultaneously using a single RF pulse (i.e., a triple resonance experiment). For ideal inversion pulses the expected frequencies are $f = \pm f_{1} \pm f_{2} $, where $f_1 = 19.0(1)$ Hz and $f_2 = 1.9(1)$ Hz are the extracted couplings between the probe spin and the two target spins (red lines). Additional frequencies emerge due to non-ideal inversion pulses (grey lines). \textbf{c}, Numerical simulation of the experiment in b) taking into account the non-ideal pulses. The obtained result matches well the experimental data and confirms that pulse errors can indeed explain the emergence of the extra observed frequencies. \textbf{d}, A quadruple-resonance experiment between a probe spin (C5) and three spectrally resolvable target spins (C2, C1, C3). See also main text Figure 3. Red lines are the theoretically expected frequencies for ideal $\pi$- pulses on the target spins, $f = \pm f_{1} \pm f_{2} \pm f_{3}$, where $f_1 = 17.17(2)$ Hz, $f_2 = 7.05(3)$ Hz and $f_3 = 3.21(4)$ Hz are the extracted couplings between the probe spin and the three target spins respectively. Grey dashed lines correspond to additional frequencies due to a failure to invert one or two of the target spins. \textbf{e}, Numerical simulation taking into account the pulse errors.}
\label{fig:multi-resonances}
\end{figure}

To address the first challenge we use two examples from our data to demonstrate that our method can extract the coupling frequencies from the complex signals in the case of multiple overlapping spins (Supplementary Fig. \ref{fig:multi-resonances}). In this scenario, our pulse sequence would essentially perform a spin echo multi-resonance experiment as the $N$ overlapping target spins will be flipped simultaneously with the probe spin. Correspondingly, the couplings between the probe spin and the $N$ target spins will be isolated. Whilst this case naturally arises when the NMR lines of multiple spins overlap, we note that the effect might more generally allow for the extraction of multiple couplings simultaneously by the application of several resonant RF pulses (Supplementary Fig. \ref{fig:multi-resonances}a) or a single spectrally broad pulse. This is a potential technique for parallelised data acquisition.

\begin{figure}[]
\centering

\includegraphics[width=0.7\textwidth]{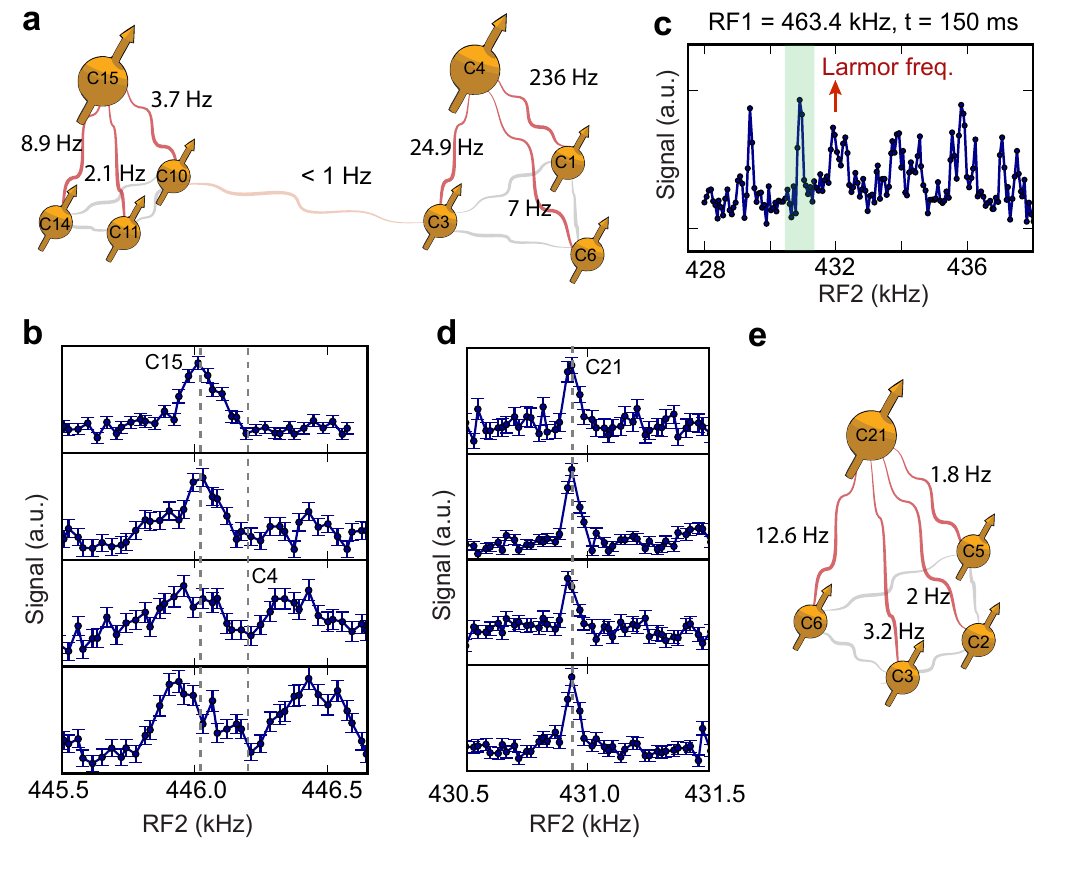}

\caption{{\bf Identifying spectrally overlapping nuclear spins.} \textbf{a}, Illustration of the basic idea:  C15 and C4 are two  nuclear spins with similar precession frequencies due to their similar coupling to the electron spin, yet we can identify and position them through their coupling to other nuclear spins. Each of these spins couples strongly to other nuclear spins in their vicinity. This connectivity reveals that there are two spins (C15 and C4) that are far apart. \textbf{b}, Double resonance frequency scan using four different probe spins (top to bottom: C14, C10, C6, C3) around a target frequency of 446 kHz. C10 and C14 are more strongly coupled to C15, while C6 and C3 are more strongly coupled to C4.  The scans for C14 and C10 show a peak at a target frequency of 446 kHz due to coupling to C15, while scans for C6 and C3 show two peaks due to a coupling to C4 (in this case C4 also strongly couples to another spin with 236 Hz which causes this splitting in its spectrum). The spectra of C15 and C4 are thus overlapping but each of them couples very differently to the 4 probe spins, revealing that there are two distinct spins. The dashed grey lines mark the centre frequencies of C15 and C4. \textbf{c}, The same idea --- i.e., nuclear spins predominantly couple to other nuclear spins in their vicinity --- can be used to uniquely identify and position nuclear spins with small couplings to the electron spin ($< 1$kHz), even though, in our system, there will generally be multiple other spins with near-identical frequencies at other locations. The shown data is a double resonance frequency scan near the bare nuclear spin Larmor frequency using C6 as a probe spin. We observe well resolved peaks potentially due to nuclear spins with relatively strong couplings to this probe spin. \textbf{d}, Double resonance scan around the shaded area in \textbf{c} ($\sim$ 1 kHz from the Larmor frequency) using 4 probe spins (top to bottom: C6, C3, C2, C5) that are spatially close to each other. \textbf{e}, The measured couplings between this target spin (C21) and the four probe spins. These measured couplings provide sufficient information to identify C21 and position it with respect to the four probe spins.}
\label{fig:overlapping spins}
\end{figure}

Supplementary Fig. \ref{fig:multi-resonances}b shows the obtained signal for a spin echo triple-resonance experiment between a probe spin (C5) and two spectrally overlapping target spins (C4, C15). The two spins have similar resonance frequencies within $200\,$Hz, and their spectra strongly overlap (Supplementary Fig. \ref{fig:overlapping spins}b). The Rabi frequency of the RF pulses is $\sim 500\,$Hz. Therefore a single resonant RF pulse on one of the target spins would flip the other target spin simultaneously.

While the obtained spectra for multi-resonance experiments are generally more complex (Supplementary Fig. \ref{fig:multi-resonances}b), the high spectral resolution in our case enables the coupling frequencies to be resolved. For the triple-resonance experiment, and assuming ideal inversion pulses,  the expected frequencies are  $f = \pm f_{1} \pm f_{2} $ where $f_1 = 19.0(1)$ Hz and $f_2 = 1.9(1)$ Hz are the extracted couplings between the probe spin and the two target spins (red lines). Due to pulse imperfections, additional frequencies also emerge (grey lines). For example, a failure in the inversion of the second target spin will create the frequency components $\pm f_{1}$ that originate from a double resonance signal between the probe and the first target spin. Similarly a failure on the inversion of the first target spin will lead to $\pm f_{2}$. We perform numerical simulations of this experiment taking into account the pulse errors (Supplementary Fig. \ref{fig:multi-resonances}c). The inversion probability of the pulses in this simulation is set to 80\%. The large infidelity in the pulses in this case is due to strong couplings of the target spins to other nearby nuclear spins, which leads to a spectrally broad signal compared to the Rabi frequency. The result of this simulation shows that pulse errors can indeed explain the emergence of the extra observed frequencies.

Supplementary Fig. \ref{fig:multi-resonances}d shows another example of a quadruple resonance experiment (Probe spin: C5, Target spins: C2, C1, C3). In this case we use 3 separate RF pulses to invert the three target spins as they have different resonance frequencies. The obtained spectrum of this measurement is even more complex, yet nevertheless the high spectral resolution makes it possible to resolve the couplings. For ideal inversion pulses, the eight theoretically expected frequencies are given by $f = \pm f_1 \pm f_2 \pm f_3$, where $f_1 = 17.17(2)$ Hz, $f_2 = 7.05(3)$ Hz and $f_3 = 3.21(4)$ Hz are the extracted couplings between the probe spin and the three target spins respectively. Additional frequencies emerge due to pulse imperfections (grey lines) as described above and confirmed by numerical simulations (Supplementary Fig. \ref{fig:multi-resonances}e).

While our scheme allows multiple couplings to be extracted from the complex obtained signal due to the high spectral resolution, this measurement alone does not yet enable the obtained couplings to be assigned to certain spins. To overcome this, we utilise the inter-connectivity between the spins in the cluster which provides enough redundancy to constrain the problem. The key idea is that each spin couples predominantly to other spins in its vicinity, and so provides a different vantage point of the cluster. So by repeating the previous measurement using different probe spins we can obtain more information to resolve these ambiguities.  Supplementary Fig. \ref{fig:overlapping spins}a,b show how we can resolve and identify the two overlapping spins, C4 and C15 shown in Supplementary Fig. \ref{fig:multi-resonances}b. 

The same idea --- nuclear spins predominantly couple to other spins in their vicinity --- enables the detection and imaging of nuclear spins with small hyperfine couplings to the electron spin ($< 5$kHz). In the system considered here, spins with small hyperfine couplings are challenging to resolve directly using the electron spin, because of a multitude of overlapping signals from spins with very similar frequencies \cite{Taminiau_PRL2012}. By using multiple spatially close nuclear spins (i.e., a sub-cluster) as probes, we can filter out the signals from remote nuclear spins at the target frequency, as we predominantly probe a certain region of the space. This allows our method to also isolate and detect, and therefore position, spins with small couplings to the electron. Supplementary Fig. \ref{fig:overlapping spins}c,d,e illustrate this concept.

\clearpage

\section{Magnetic field alignment by thermal echo}\label{sec:bfieldalignment}

\begin{figure}
\centering
\includegraphics[width=0.9\textwidth]{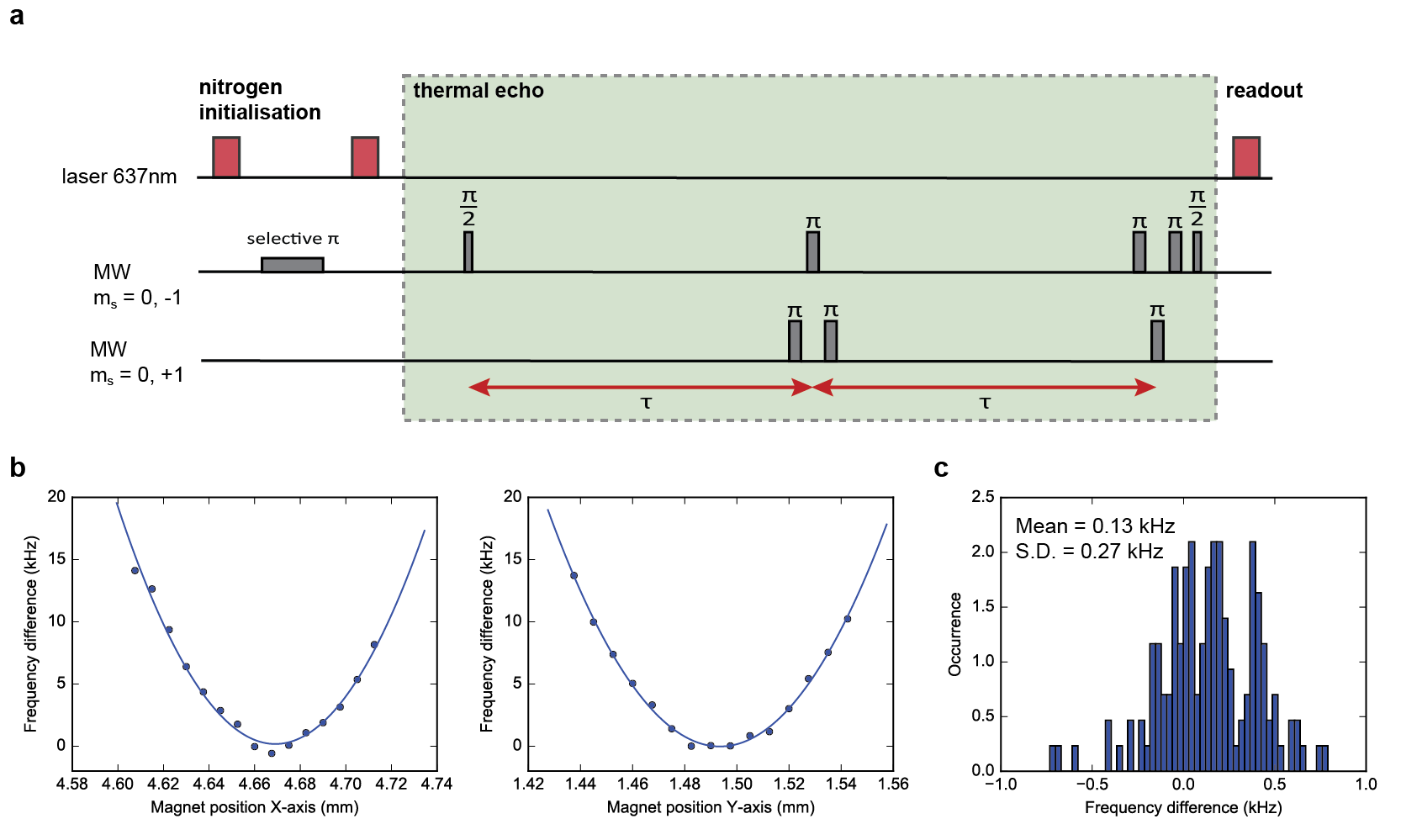}

\caption{{\bf Aligning the magnetic field using a thermal echo sequence} \textbf{a,} Pulse sequence for the thermal echo measurement. Laser pulses at $\sim 637\,$nm are used to initialise and read out the electron spin (top line), while microwave (MW) pulses at $\sim 1.7\,$GHz ($\sim 4.0\,$GHz) are used to manipulate the electron spin state on the $m_s = 0 \leftrightarrow m_s = -1$ ($m_s = +1$) transition. \textbf{b,} We scan the magnet position in two orthogonal directions (first along $x$ and then along $y$), performing a thermal echo measurement at each position. The obtained frequency differences ($= f - f_\text{min}$, where $f_\text{min} = 2.877652 \,$GHz is the minimum measured frequency) are plotted versus the magnet position on both axes. We fit the obtained data to a parabolic function to find the optimum position in x and y.  \textbf{c,} Placing the magnet at the optimum position and repeating the measurement 200 times (over a 10-hour period). The obtained average frequency difference is $0.13$ kHz, with a standard deviation of 0.27 kHz, which is consistent with the estimated statistical measurement error.}
\label{fig:thermal_echo}
\end{figure}

We align the magnetic field along the NV axis, (the $z$-axis; $B_z \sim 403\,$G). A non-zero perpendicular component, $B_\perp$, can introduce additional electron-nuclear spin mixing and thus modify the measured nuclear-nuclear couplings (see section \ref{sec:corrections}). Therefore, a precise alignment of the field is required to obtain a faithful reconstruction of the structure of the spin cluster. Here we exploit a `thermal' echo sequence, which has been previously introduced for measuring temperature \cite{Toyli8417}, to align the magnet field with uncertainty below $0.07$ degrees or $B_\perp = 0.5\,$G. 

The thermal echo sequence is as follows (see Supplementary Fig. \ref{fig:thermal_echo}(a)). First, the nitrogen spin is initialised in $\ket{m_I = 0}$ using measurement based initialisation \cite{Robledo_Nature2011} and the electron spin is prepared in $\ket{m_s = 0}$. Following this, a rotation $R(\pi/2,0)_{-1}$ brings the electron to the equal superposition state $(\ket{m_s = 0} + \ket{m_s = -1})/\sqrt{2}$, where we have used the notation $R(\theta,\phi)_{j}$ to indicate a rotation by angle $\theta$ around an axis $\cos(\phi)\hat{x} + \sin(\phi)\hat{y}$ between states $\ket{m_s = 0}$ and $\ket{m_s = j}$. The state is then allowed to evolve for a time $\tau$, after which an effective $\pi$-pulse between $\ket{m_s = -1}$ and $\ket{m_s = +1}$ is performed, using the sequence $R(\pi,0)_{+1} - R(\pi,0)_{-1} - R(\pi,0)_{+1}$. After a second free evolution time $\tau$, a second effective $\pi$-pulse is performed (using the sequence $R(\pi,0)_{-1} - R(\pi,0)_{+1} - R(\pi,0)_{-1}$), followed by a final rotation $R(\pi/2,0)_{-1}$. Since the energies of the states $\ket{m_s = \pm 1}$ are shifted by equal and opposite amounts by Hamiltonian terms proportional to $S_z$, the effects of such terms are cancelled for the thermal echo sequence, similar to a standard spin echo. However, for Hamiltonian terms that shift the energies of $\ket{m_s = \pm 1}$ in the same way, such as the zero-field splitting and magnetic fields perpendicular to $z$, the effects do not cancel. Therefore, the thermal echo cancels the effect of the biggest noise source (magnetic field noise along $z$ from the surrounding spin bath), while remaining sensitive to smaller shifts caused by the perpendicular magnetic field components, allowing them to be measured more accurately. By sweeping $\tau$, a measurement of the frequency difference $f_\text{TE} = (f_+ + f_-)/2$ is obtained, which is minimised when $B_\perp = 0$.

We scan the magnet position in two orthogonal directions (first along $x$ and then along $y$) and then perform the thermal echo measurement. Supplementary Fig. \ref{fig:thermal_echo}b shows the obtained frequency. We fit the obtained data to a parabolic function to find the optimum position in $x$ and $y$. Afterwards, we place the magnet at the optimum position and repeat the measurement several times over a $10$-hour period (Supplementary Fig. \ref{fig:thermal_echo}c). The obtained average frequency difference is $0.13$ kHz with a standard deviation of 0.27 kHz, which is consistent with the thermal echo measurement uncertainty. Therefore, the total uncertainty in our magnet alignment is 0.4 kHz which corresponds to a perpendicular field of 0.5 Gauss or a misalignment angle of 0.07 degrees.

\section{Accounting for electron-mediated couplings}\label{sec:corrections}

The nuclear-nuclear couplings measured by a double-echo sequence can be modified due to the presence of the electron spin and a misaligned magnetic field. To understand these effects, we can use perturbation theory. In refs. \cite{Childress_Science2006, Zhao_NatureNat2011}, shifts in the nuclear-nuclear couplings are calculated by only considering the interaction between the electron and $^{13}$C spins in the perturbation. However, for the weak electron-nuclear couplings considered in this work, modifications due to the non-secular nuclear-nuclear interactions combined with the electron-nuclear interactions can also give a significant correction, and should therefore also be included in the perturbation. We will consider the Hamiltonian describing the spin-1 electron plus two spin-1/2 $^{13}$C spins, given by

\begin{equation}\label{eq:Hfull}\begin{split}
    H &= H_e + H_c + H_{ec} + H_{cc}, \\
    H_e &= \Delta_\text{ZFS} S_z^2 + \gamma_e (B_x S_x + B_y S_y + B_z S_z), \\
    H_c &= \gamma_c \vec{B}\cdot(\vec{I}^{(1)} + \vec{I}^{(2)}), \\
    H_{ec} &= \vec{S} \cdot {\bf A}^{(1)} \cdot \vec{I}^{(1)} + \vec{S} \cdot {\bf A}^{(2)} \cdot \vec{I}^{(2)}, \\
    H_{cc} &= \vec{I}^{(1)} \cdot {\bf C} \cdot \vec{I}^{(2)},
\end{split}\end{equation}
where $\Delta_\text{ZFS}$ is the electron zero field splitting, $\gamma_e$ ($\gamma_c$) is the electron ($^{13}$C) gyromagnetic ratio, $\vec{B} = (B_x, B_y, B_z)$ is the magnetic field vector, $\vec{S} = (S_x, S_y, S_z)$ are the electron spin operators, $\vec{I}^{(j)} = (I_x^{(j)}, I_y^{(j)}, I_z^{(j)})$ are the $^{13}$C spin operators for spin $j$, and ${\bf A}^{(j)}$ and ${\bf C}$ are the hyperfine tensors describing the electron-nuclear and nuclear-nuclear interactions respectively, with components $A_{\alpha\beta}^{(j)}$ and $C_{\alpha\beta}$ for $\alpha, \beta \in \{x,y,z\}$.

In our experiments we apply a strong magnetic field along the $z$-axis ($B_z \sim 403\,$G), and align the field such that $B_x, B_y \approx 0$. The dominant energy scales are then given by the terms $\Delta_\text{ZFS}S_z^2$, $\gamma_e B_z S_z$ and $\gamma_c B_z I_z^{(j)}$. Therefore, we will take terms that commute with $S_z^2$, $S_z$, $I_z^{(1)}$ and $I_z^{(2)}$ as the unperturbed Hamiltonian $H_0$. This gives

\begin{equation}
    H_0 = \Delta_\text{ZFS} S_z^2 + \gamma_e B_z S_z + \gamma_c B_z (I_z^{(1)} + I_z^{(2)}) + A_{zz}^{(1)} S_z I_z^{(1)} + A_{zz}^{(2)} S_z I_z^{(2)} + C_{zz}I_z^{(1)}I_z^{(2)}.
\end{equation}
The eigenstates of $H_0$ are $|m_s, m_I^{(1)}, m_I^{(2)}\rangle$, where $m_s \in \{+1, 0, -1\}$ are the eigenvalues of $S_z$ and $m_I^{(j)} \in \{+\frac{1}{2}, -\frac{1}{2}\}$ are the eigenvalues of $I_z^{(j)}$. The eigenvalues of $H_0$ are then

\begin{equation}\begin{split}
    \lambda_0(m_s, m_I^{(1)}, m_I^{(2)}) = &m_s^2\Delta_\text{ZFS} + m_s\gamma_e B_z +  (m_I^{(1)} + m_I^{(2)})\gamma_cB_z \\
    &+ m_s m_I^{(1)}A_{zz}^{(1)} + m_s m_I^{(2)}A_{zz}^{(2)} + m_I^{(1)} m_I^{(2)}C_{zz}.
\end{split}\end{equation}
A double-resonance measurement gives an oscillating signal at frequency \cite{slichter}

\begin{equation}
    f_\text{DE}(m_s)= \frac{1}{4\pi}\left|\lambda\left(m_s, +\frac{1}{2}, +\frac{1}{2}\right) + \lambda\left(m_s, -\frac{1}{2}, -\frac{1}{2}\right) - \lambda\left(m_s, +\frac{1}{2}, -\frac{1}{2}\right) -\lambda\left(m_s, -\frac{1}{2}, +\frac{1}{2}\right)\right|.
\end{equation}
If we take the zeroth order approximation, such that $\lambda(m_s, m_I^{(1)}, m_I^{(2)}) \approx \lambda_0(m_s, m_I^{(1)}, m_I^{(2)})$, we obtain

\begin{equation}
    f_\text{DE} \approx \frac{1}{4\pi}|C_{zz}|,
\end{equation}
which holds for all electron spin projections. To zeroth order, the double resonance measurement gives the $zz$ dipolar coupling between the two $^{13}$C spins, as expected. 

We can now calculate corrections to the nuclear-nuclear couplings due to the other Hamiltonian terms using perturbation theory. Defining the remaining terms in the Hamiltonian as $V = H - H_0$, up to second order the corrected eigenvalues are given by \cite{Sakurai:1167961}

\begin{equation}\label{eq:perturbation}
    \lambda(\psi_n) \approx \lambda_0(\psi_n) + \bra{\psi_n}V\ket{\psi_n} + \sum_{k\neq n}\frac{|\bra{\psi_k}V\ket{\psi_n}|^2}{\lambda_0(\psi_n) - \lambda_0(\psi_k)},
\end{equation}
where $\ket{\psi_n}$ are the eigenstates of $H_0$. Since $V$ contains no diagonal matrix elements, $\bra{\psi_n}V\ket{\psi_n} = 0$ for all $\ket{\psi_n}$. We will restrict our analysis to the $m_s = \pm 1$ subspace, as only transitions within this subspace are measured for the experiments described in this manuscript and for the $m_s = 0$ subspace some eigenstates are degenerate, making the analysis more complex. We find three sets of correction terms, allowing us to write the corrected double resonance frequency up to second order as 

\begin{equation}\label{eq:DEms}
    f_\text{DE}(m_s = \pm 1) \approx \frac{1}{4\pi}|C_{zz} + \Delta\lambda_1(m_s) + \Delta\lambda_2(m_s) + \Delta\lambda_3(m_s) |.
\end{equation}
The first correction term $\Delta\lambda_1$ describes a correction to the nuclear-nuclear coupling that is only dependent on the interaction between each $^{13}$C spin and the electron spin. This term is equal to the correction term derived in refs. \cite{Childress_Science2006, Zhao_NatureNat2011}. With the approximation that $|\gamma_cB_z|, |A_{zz}^{(j)}|, |C_{zz}| \ll |\Delta_\text{ZFS} + m_s\gamma_e B_z|$, we obtain

\begin{equation}\label{eq:lambda1}
    \Delta\lambda_1(m_s = \pm 1) \approx \frac{A_{zx}^{(1)}A_{zx}^{(2)} + A_{zy}^{(1)}A_{zy}^{(2)}}{\Delta_\text{ZFS} + m_s\gamma_e B_z}.
\end{equation}
The second correction term $\Delta\lambda_2$ describes a correction that depends on both the nuclear-electron and the nuclear-nuclear interactions. With the approximation that $|C_{zz}| \ll |\gamma_cB_z|$, and making a Taylor expansion up to first order in $A_{zz}^{(j)}/\gamma_c B_z$, we find

\begin{figure}
\centering

\includegraphics[width=0.9\textwidth]{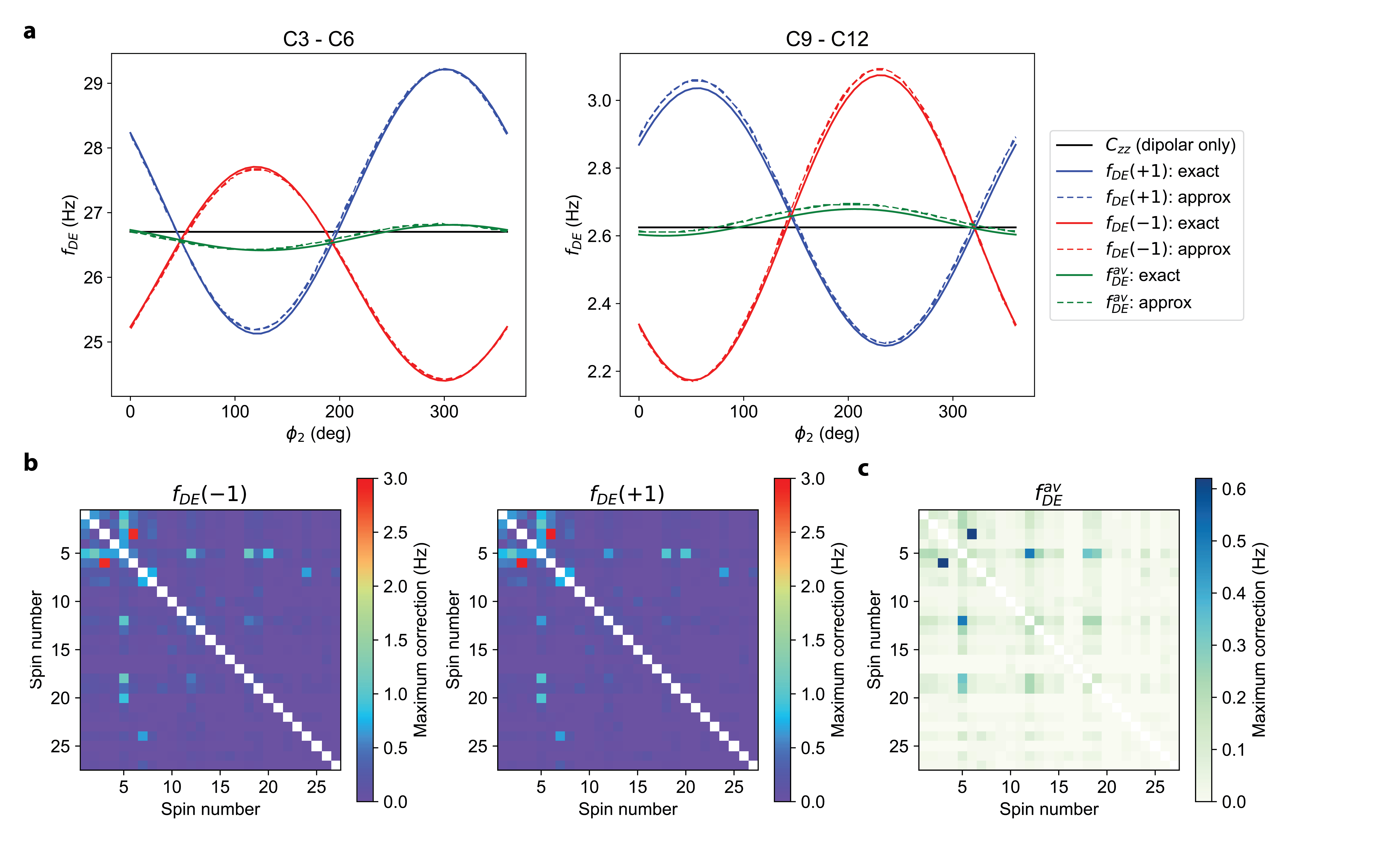}

\caption{{\bf Corrections to the measured couplings} \textbf{a,} Two examples of modified nuclear-nuclear couplings due to electron mediated effects for exact numerical solutions (solid lines) and approximate solutions using equations \ref{eq:DEms} and \ref{eq:DEav} (dashed lines). The hyperfine parameters are taken from Supplementary Table \ref{table:couplingav}. The radial magnetic field component $B_\perp$ is set to zero in these examples, but similar results are seen for small values $< 1\,$G as measured experimentally (see section \ref{sec:bfieldalignment}). The nuclear-nuclear hyperfine tensors are calculated from the coordinates given in Supplementary Table \ref{table:coordinates} and the unknown angles $\varphi_j$ are set to 0 for the first spin and varied for the second spin (other values for the first spin also give similar agreement). \textbf{b,} Maximum corrections to the measured nuclear-nuclear couplings between each of the 27 spins for the $m_s = \pm 1$ spin projections, $f_\text{DE}(\pm1)$, calculated from equation \ref{eq:DEms}. For each coupling the correction is maximised over the unknown parameters $0 < \varphi_j < 2\pi$, $0 < B_\perp < 1\,$G and $0 < \theta <2\pi$. \textbf{c,} The same correction matrix as in (b) for the averaged coupling $f_\text{DE}^\text{av}$ (equation \ref{eq:DEav}). }
\label{fig:examples}
\end{figure}

\begin{equation}
    \Delta\lambda_2(m_s = \pm 1) \approx m_s \Delta\lambda_2^{(0)} + \Delta\lambda_2^{(1)}
\end{equation}
where $m_s\Delta\lambda_2^{(0)}$ and $\Delta\lambda_2^{(1)}$ are the zeroth and first order terms in the Taylor expansion respectively, given by
\begin{equation}\label{eq:lambda2}\begin{split}
    \Delta\lambda_2^{(0)} &= \frac{(A_{zx}^{(1)} + A_{zx}^{(2)}) C_{zx} + (A_{zy}^{(1)} + A_{zy}^{(2)}) C_{zy}}{\gamma_c B_z} \\
    \Delta\lambda_2^{(1)} &= - \sum_{j=1}^2 \frac{(A_{zx}^{(j)}C_{zx} + A_{zy}^{(j)}C_{zy})A_{zz}^{(j)}}{\gamma_c^2 B_z^2}.
\end{split}\end{equation}
Lastly, there is a correction that depends on the perpendicular magnetic field. With the approximation that $|C_{zz}| \ll |\gamma_cB_z|$, and making a Taylor expansion up to first order in $A_{zz}^{(j)}/\gamma_c B_z$, we obtain

\begin{equation}
    \Delta\lambda_3(m_s = \pm 1) \approx \Delta\lambda_3^{(0)} + m_s\Delta\lambda_3^{(1)}
\end{equation}
where $\Delta\lambda_3^{(0)}$ and $m_s\Delta\lambda_3^{(1)}$ are the zeroth and first order terms in the Taylor expansion respectively, given by
\begin{equation}\label{eq:lambda3}\begin{split}
    \Delta\lambda_3^{(0)} &= \frac{2(B_x C_{zx} + B_y C_{zy})}{B_z} \\
    \Delta\lambda_3^{(1)} &= \frac{(A_{zz}^{(1)} + A_{zz}^{(2)})(B_x C_{zx} + B_y C_{zy})}{\gamma_cB_z^2}.
\end{split}\end{equation}
We can now calculate the average frequency for a double resonance measurement on both $m_s = \pm 1$ transitions. We find

\begin{equation}\label{eq:DEav}\begin{split}
    f_\text{DE}^\text{av} &= \frac{1}{2}\big(f_\text{DE}(+1) + f_\text{DE}(-1)\big) \\
    &= \frac{1}{4\pi}\left|C_{zz} + \frac{\Delta\lambda_1(+1) + \Delta\lambda_1(-1)}{2} + \Delta\lambda_2^{(1)} + \Delta\lambda_3^{(0)}\right|,
\end{split}\end{equation}
where in the last line we have assumed that $|\Delta\lambda_1(m_s) + \Delta\lambda_2(m_s) + \Delta\lambda_3(m_s)| < |C_{zz}|$. It can be seen that the terms $m_s\Delta\lambda_2^{(0)}$ and $m_s\Delta\lambda_3^{(1)}$ cancel when taking the average double resonance frequency.

To check the validity of the approximate solutions, we can compare to exact numerical solutions of the full Hamiltonian (equation \ref{eq:Hfull}) for some example cases. Since we only measure the magnitude of the perpendicular hyperfine couplings $A_\perp^{(j)} = \sqrt{\left(A_{zx}^{(j)}\right)^2 + \left(A_{zy}^{(j)}\right)^2}$, we will parameterise the couplings as $A_{zx}^{(j)} = A_\perp^{(j)}\cos(\varphi_j)$ and $A_{zy}^{(j)} = A_\perp^{(j)}\sin(\varphi_j)$, where $\varphi_j$ is the unknown azimuthal angle of the electron-nuclear hyperfine coupling for spin $j$. Additionally, we can parameterise the magnetic field in the $x-y$ plane as $B_x = B_\perp\cos(\theta)$ and $B_y = B_\perp\sin(\theta)$. From section \ref{sec:bfieldalignment}, we can infer that $B_\perp < 1\,$G, while no information about $\theta$ is known. Supplementary Fig. \ref{fig:examples}(a) shows a comparison between the approximate and exact solutions for two example cases based on parameters measured in this experiment. We can also estimate the magnitude of each correction term using the values for the parameters used in the experiment. Taking the coordinates of the 27 spin cluster (Supplementary Table \ref{table:coordinates}), for each possible spin pair we can maximise equations \ref{eq:DEms} and \ref{eq:DEav} over the unknown angles $\varphi_j$ and the magnetic field parameters $B_\perp$ and $\theta$. Matrix plots of estimated maximum values are shown in Supplementary Fig. \ref{fig:examples}(b,c). Averaging over the unknown angles $\varphi_j$, the average correction over all spins is $0.04\,$Hz for $m_s = \pm 1$ and $0.01\,$Hz for the average, while the maximum over all $\varphi_j$ and spins is $2.6\,$Hz ($3.1\,$Hz) for $m_s = -1$ ($m_s = +1$) and $0.55\,$Hz for the average. The corrections for $m_s = \pm 1$ are therefore much greater than the averaged values, thereby showing that the measured couplings are closer to the dipolar values when taking the average of the $m_s = \pm 1$ measurements.

\section{Obtaining the structure of the nuclear spin cluster}

We would like to find the relative coordinates of $M$ nuclear spins: $\{x_i, y_i, z_i\}$, $i = 1, ..., M$. Each nuclear spin is pairwise coupled to every other spin with $zz$ coupling constants $C_{ij}(x_i,y_i,z_i,x_j,y_j,z_j)$, which, assuming point-dipole coupling, are related to the coordinates by the set of equations

\begin{equation}\label{eq:hyperfine}
    C_{ij} = \frac{\alpha_{ij}}{\Delta r_{ij}^3}\left(\frac{3(z_j - z_i)^2}{\Delta r_{ij}^2}-1\right),
\end{equation}
where $\Delta r_{ij} = \sqrt{ (x_j - x_i)^2 + (y_j - y_i)^2 + (z_j - z_i)^2 }$, $\alpha_{ij} = \mu_0 \gamma_i \gamma_j \hbar/4\pi$, $\mu_0$ is the permeability of free space, $\gamma_i$ is the gyromagnetic ratio of nuclear spin $i$ and $\hbar$ is the reduced Planck constant. A double resonance measurement performed on two spins $i$ and $j$ gives a signal oscillating at frequency $f_{ij}$, which is approximately related to the magnitude of the coupling as $f_{ij} \approx |C_{ij}|/4\pi$. Therefore, we define the residual for each coupling as $\Delta f_{ij} \equiv f_{ij} - |C_{ij}|/4\pi$. There are $3M$ coordinates and $M(M-1)/2$ coupling constants. Since we are interested in only the relative coordinates, we can fix the first spin to be at the origin: $\{x_1, y_1, z_1\} = \{0,0,0\}$. Additionally, since there is a rotational symmetry in the $x-y$ plane, we can also set one of the $x,y$ coordinates of the second spin to zero. The number of free coordinates is therefore $3M - 4$, and consequently to achieve more measurements than free parameters, as required for the problem to be overdetermined, we require $N \geq 6$ (for $N= 6$: $3M - 4 = 14$, $M(M-1)/2 = 15$). The best fit solution is then given by a set of parameters  $\{x_i, y_i, z_i\}$ that minimise the sum of squares, defined as 

\begin{equation}\label{eq:sumsq}
    \xi \equiv \sum_{i = 1}^M \sum_{j = 1}^i |\Delta f_{ij}|^2.
\end{equation}
 
As described in section \ref{sec:corrections}, the largest of the corrections to the couplings (due to the presence of the electron spin) are cancelled when taking the average of the frequencies measured for the $m_s = \pm 1$ electron spin projections. Despite this, the measurement uncertainties can be smaller than the corrections (measurement uncertainties are typically $< 0.1\,$Hz, while the corrections could be up to $\sim 0.6\,$Hz in the worst case). Since the corrections depend on unknown parameters (such as the azimuthal angles between the electron wavefunction and the $^{13}$C spins), we consider them as an additional source of uncertainty in this work. Techniques to measure these unknown parameters have been developed in refs. \cite{Zopes_NatComm2018, Zopes_PRL2018}. Combining our methods with those techniques could in the future be used for precision spectroscopy on these coupled electron-nuclear-nuclear systems.

A common method to minimise $\xi$ is to use a fitting algorithm such as least-squares minimisation. However, with no a priori information regarding the structure, finding an initial guess where the fit will converge to the global minimum becomes difficult. By testing with randomly generated clusters within a $1\,$nm$^3$ volume, we found that the initial guess should be within approximately $0.5\,$\AA\ for each $x,y,z$ component for each spin in order for the fit to converge to the true solution. For 27 spins, this corresponds to an intractable $\sim 10^{100}$ required initial guesses to cover the entire search space.

A solution is to build up the configuration on a three-dimensional grid by adding spins one-by-one, while tracking all solutions within an error tolerance. We do this using two different methods. In the first method, we use a priori knowledge of the diamond lattice to constrain the possible spin positions (section \ref{sec:diamondlattice}). This is efficient for nuclear spins in diamond, but cannot be used for an arbitrary spin system. The second method uses a finer cubic lattice (section \ref{sec:cubiclattice}), which is more computationally intensive, but is general to any spin system measured using the methods described in this work. We can also use these solutions as an initial guess for a fit of the spatial coordinates to the measured couplings.

\subsection{Positioning spins using the diamond lattice}\label{sec:diamondlattice}

Since we know that the $^{13}$C spins identified in this work are located at points on the diamond lattice, it is efficient to constrain the spatial coordinates to possible lattice sites. The procedure used to find the configuration is as follows. Firstly, the coordinates of a diamond lattice are generated with $2 N_L + 1$ points along each of the $[0,1,1]$, $[1,0,1]$ and $[1,1,0]$ crystal axes, where $N_L$ is an integer, spaced by $a_0/\sqrt{2}$, where $a_0 = 3.5668\,$\AA\ is the diamond lattice constant \cite{diamond_lattice}. An additional lattice site for each point is then added at a displacement $\frac{a_0}{4}[1,1,1]$, and the lattice is oriented such that the $[1,1,1]$ direction is parallel to the $z$-axis with the origin at the centre. The total number of lattice sites is therefore $2(2 N_L + 1)^3$. For each lattice site, the coupling to a spin at the origin is calculated using equation \ref{eq:hyperfine} and stored in a lookup table along with the corresponding spatial vector between the two coordinates. Starting from an initial spin placed at the origin, each spin is placed in turn at a lattice site by choosing the strongest measured coupling to any spin that has already been placed and finding all corresponding possible vectors from the lookup table. Vectors from the lookup table are selected if the theoretically calculated coupling satisfies $\Delta f_{ij} < \mathcal{T}$, where $\mathcal{T}$ is a tolerance that is chosen to ensure that all promising configurations are included while maintaining reasonable computation time. 

Next, for each possible configuration all theoretically calculated pairwise couplings are compared with the experimentally measured values and are also required to satisfy $\Delta f_{ij} < \mathcal{T}$. Configurations that meet the criteria for all couplings are kept and the procedure is then repeated for the next spin. For the second spin added only one of the six possible C$_\text{3v}$-symmetric configurations is kept. Once this procedure has been performed for all spins in the measured cluster, if multiple configurations have been found, the best solution can be determined by minimising the sum of squares $\xi$ (equation \ref{eq:sumsq}). We use $N = 1$ echo pulses for most coupling measurements in Supplementary Table \ref{table:couplingav}. For some weak couplings, no oscillation could be observed within the decay time for $N=1$. Such weak couplings can be measured more accurately using multiple pulses (e.g. Fig. 1f of the main text) or by a low-noise comparison to the Hahn echo. However, since the precise values of such weak couplings have a small effect on the obtained solution and due to time constraints, we did not perform such measurements for all couplings. Couplings that have been determined to be weak, but for which no precise value was established, are marked as $< 1\,$Hz (Supplementary Tables \ref{table:couplingmsm1}-\ref{table:couplingav}). For these measurements, we use a value of $0.5\,$Hz in the analysis. 

\begin{figure}
\centering

\includegraphics[width=1\textwidth]{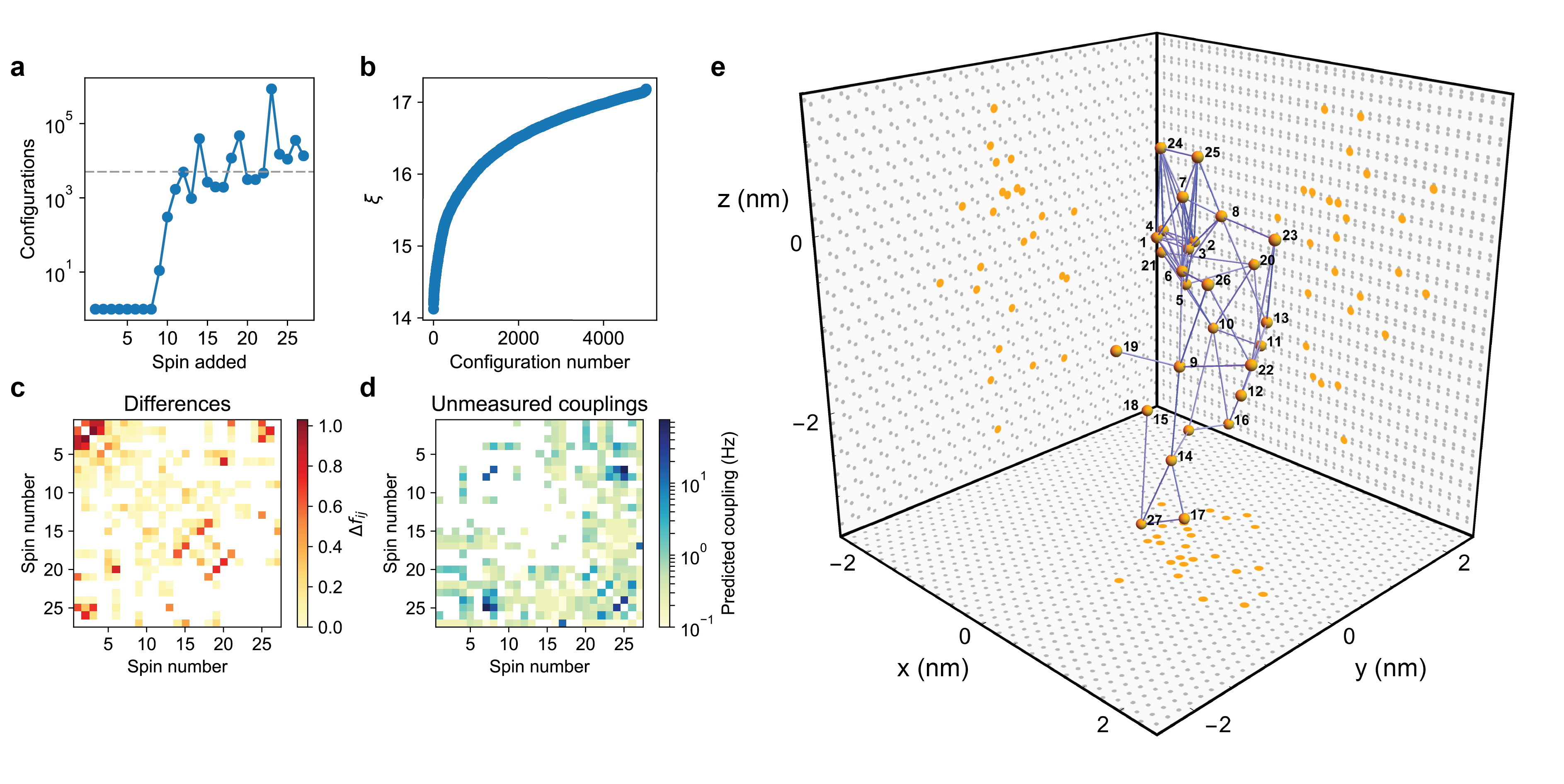}

\caption{{\bf Obtaining the structure using the diamond lattice.} \textbf{a,} Total number of configurations found after each spin added (log scale). The dashed grey line indicates the cutoff point set at $X_\text{cutoff} = 5000$. If more than $X_\text{cutoff}$ solutions are found for a given spin (the point is above the dashed line), only the best $X_\text{cutoff}$ solutions are kept for the next spin. \textbf{b,} Sum of squares $\xi$ calculated from equation \ref{eq:sumsq} for all configurations found for 27 spins, arranged in order from best to worst. \textbf{c,} Matrix plot showing the residuals $\Delta f_{ij}$ for the configuration with the lowest $\xi$. White squares indicate unmeasured couplings. \textbf{d,} Matrix plot showing the predicted values of $|C_{ij}|/4\pi$ for couplings that were not measured. White squares indicate measured couplings. \textbf{e,} 3D plot of the 27 spin configuration with the lowest $\xi$ value. The three sides show the $x-y$, $x-z$ and $y-z$ projections of the spins (yellow points) and the diamond lattice coordinates (grey points). Blue lines represent couplings greater than $3\,$Hz.}
\label{fig:positioning}
\end{figure}

For the solution presented in Fig. 4 of the main text and also in Supplementary Fig. \ref{fig:positioning}, the diamond lattice coordinates were generated using $N_L = 11$, corresponding to approximately $24 \times 10^3$ lattice sites and a volume of around $120\,$nm$^3$. The tolerance was chosen to be $\mathcal{T} = 1.1\,$Hz, so that the best solution is among the configurations identified with a high probability. For a small number of measured couplings, the double resonance measurement was only performed for one of the $m_s = \pm 1$ electron spin projections, and therefore the higher order corrections (section \ref{sec:corrections}) can be larger. For these couplings we allowed a higher tolerance of $\mathcal{T}_\text{single} = 3\,$Hz. In addition, a cutoff limit of $X_\text{cutoff} = 5000$ was set for the maximum number of solutions that are carried over to the next spin. If the number of solutions found exceeds $X_\text{cutoff}$, the solutions are ordered from best to worst according to their $\xi$ value and only the first $X_\text{cutoff}$ are saved.

For the first 8 spins, which form a strongly coupled sub-cluster, only one solution is found for each spin added (Supplementary Fig. \ref{fig:positioning}a). After this, more weakly coupled spins are added, which can give multiple configurations that satisfy the criteria. The final 27 spin solution is given by the solution with the lowest $\xi$ (Supplementary Fig. \ref{fig:positioning}b). Coordinates for this configuration are given in Supplementary Table \ref{table:coordinates} and the matrix of differences $\Delta f_{ij}$, the unmeasured couplings and a 3D plot are shown in Supplementary Fig \ref{fig:positioning}c,d,e. Two identified sub-clusters are shown in greater detail in Supplementary Fig. \ref{fig:subclusters}. 19 spins have the same coordinates in all 5000 configurations found, while the remaining 8 spins (18, 19, 20, 23, 24, 25, 26 and 27) have multiple positions within the range of identified solutions. The routine took approximately $6\,$hours on a desktop PC.

Using the solution found for the spatial coordinates of the 27 spin cluster, we can calculate the number of expected spins in the total volume to estimate the fraction of spins in this region we have identified. Defining a rectangular box around the cluster defined by the minimum and maximum $x$, $y$ and $z$ coordinates over all spins, we get a volume of $2 \times 2 \times 4 = 16\,$nm$^3$. This volume contains approximately $2900$ lattice sites. With a natural $^{13}$C concentration of $1.1\%$, we therefore expect approximately 32 spins in this volume, consistent with the 27 spins identified. 

\begin{figure}
\centering

\includegraphics[width=0.8\textwidth]{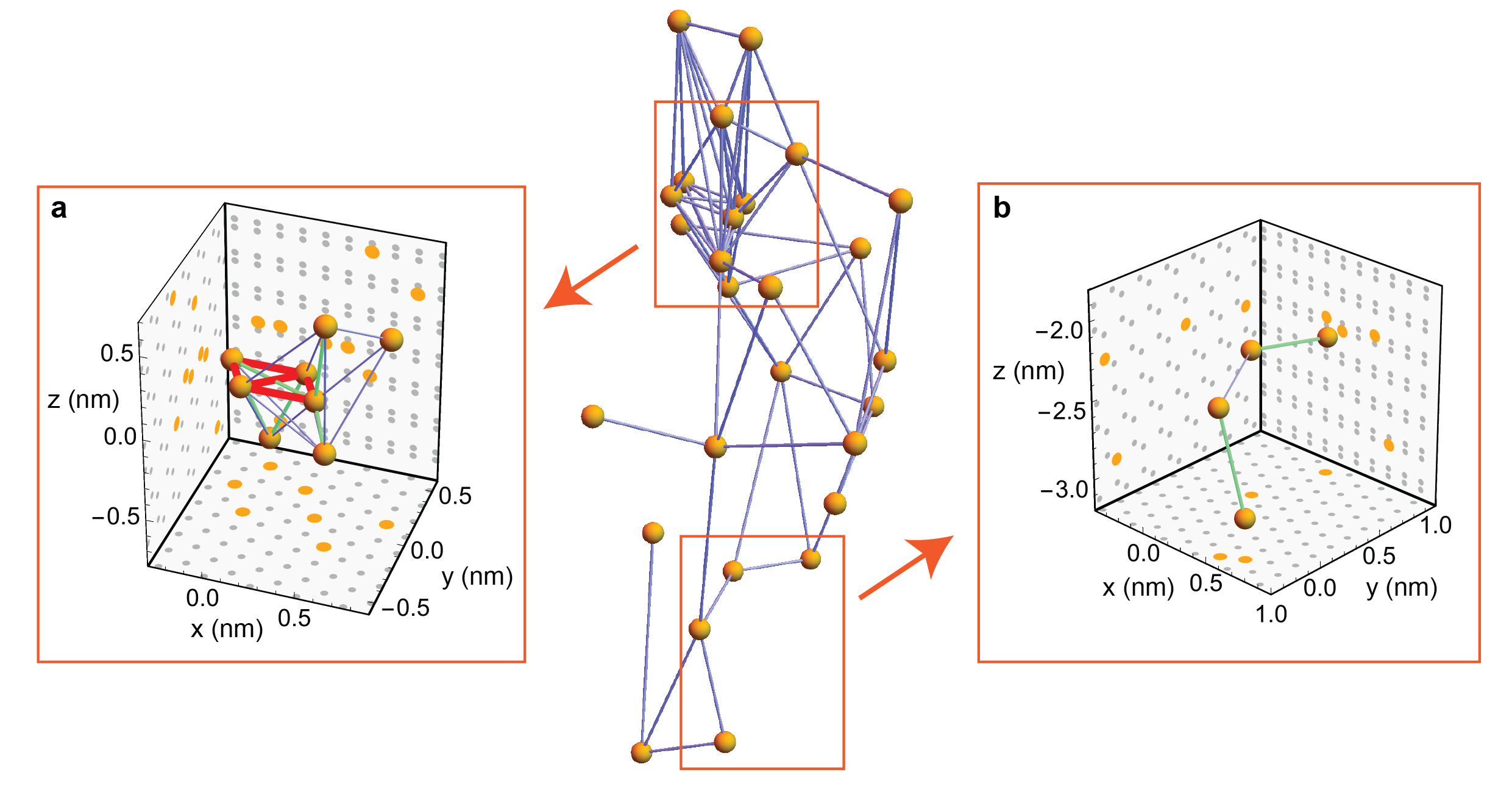}

\caption{\textbf{Strongly coupled subclusters.} 3D plots showing the structure of two strongly coupled subclusters (orange panels) within the larger cluster (shown in the centre). The positions of the subclusters within the larger cluster are marked by the orange boxes and arrows. Ramsey measurements performed on spins within these subclusters show strong beatings within their $T_2^*$ dephasing time (see Supplementary Fig. \ref{fig:Ramseys}). Panel a) shows the 8 spin subcluster, while panel b) shows the 4 spin subcluster. Couplings above $3\,$Hz are marked blue, above $20\,$Hz green and above $50\,$Hz red.  }
\label{fig:subclusters}
\end{figure}

\subsection{Least-squares minimisation}\label{sec:fit}

In addition to finding to structure of the cluster using the diamond lattice, we can further use the obtained solution as an initial guess for a least-squares minimisation routine. To properly constrain the fit, all three coordinates of spin 1 and additionally the $y$ coordinate of spin 2 are fixed to zero in order to constrain to only relative positions and to break the rotational symmetry in the $x-y$ plane. The solution used for the initial guess is therefore rotated around the $z$-axis by an angle $\phi = -49.1\,$deg so that the rotated coordinate $y_2' = 0$. To quantify the difference between the initial and fitted solutions, we can define the distance from the initial guess for each spin as $\Delta r_i = \sqrt{\Delta x_i^2 + \Delta y_i^2 + \Delta z_i^2}$, where $\{\Delta x_i, \Delta y_i, \Delta z_i \}$ are the differences between the fitted coordinates and the initial guess coordinates (from the diamond lattice solution) for spin $i$. A bar plot of $\Delta r_i$ for each spin in the fitted 27 spin solution, along with the associated fit uncertainties, is shown in Supplementary Fig. \ref{fig:fitresult}a. The average distance between the solutions is $\frac{1}{M}\sum_{i}^M \Delta r_i = 0.46$\AA. Since the measurement uncertainties are not well known due to the electron-mediated coupling corrections (section \ref{sec:corrections}), we calculate the fit parameter errors based only on the variance of the residuals. It can be seen that for the majority of spins, the fitted coordinates and uncertainties are within one diamond bond length from the configuration entered as the initial guess.

\begin{figure}
\centering

\includegraphics[width=0.7\textwidth]{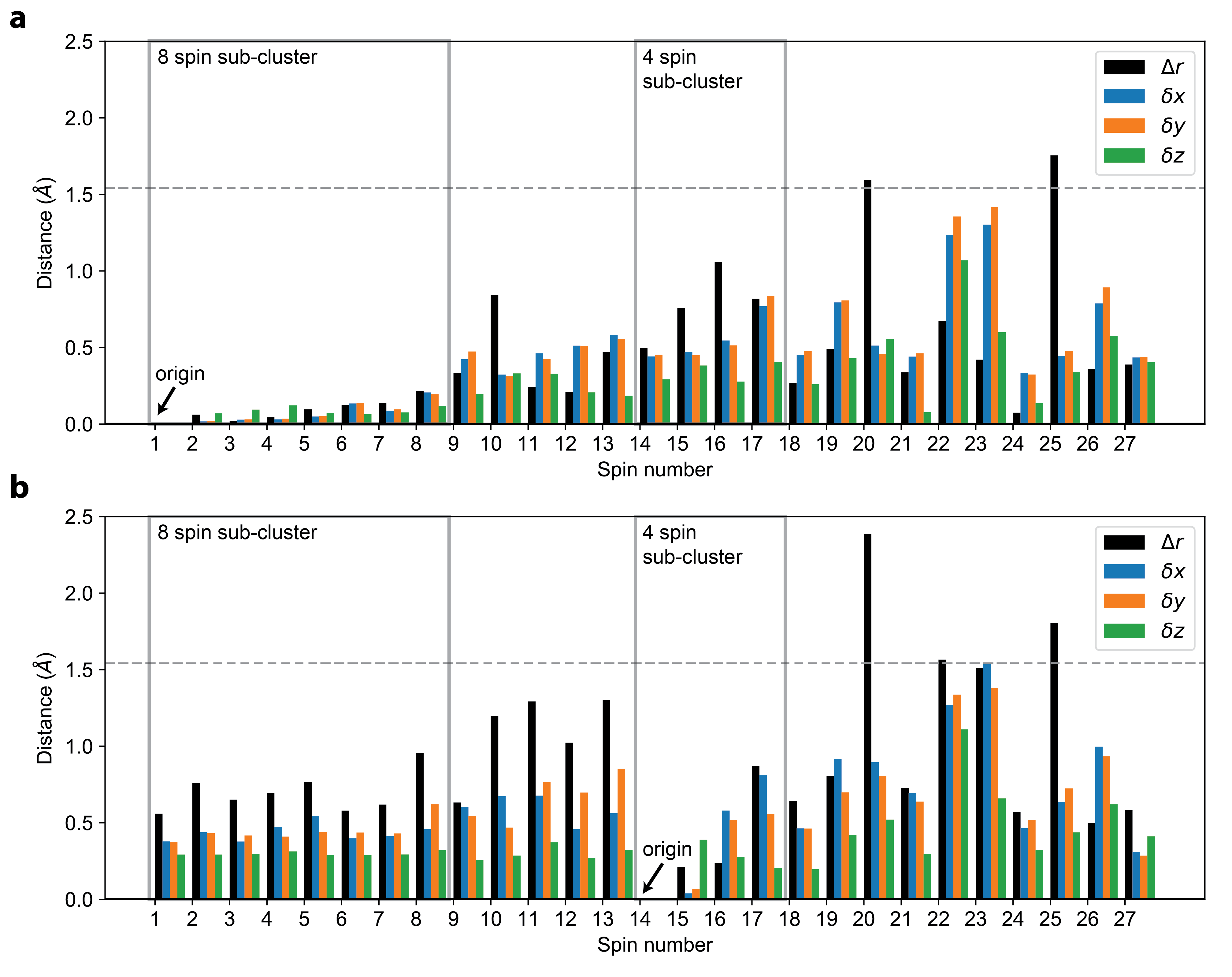}

\caption{{\bf Fitted solution for 27 spins.} Bar plots of $\Delta r$ for each spin (black), defined as the magnitude of the distance between the fitted coordinates and the initial guess, as well as fit errors $\delta x$ (blue), $\delta y$ (orange) and $\delta z$ (green). The initial guess was found using the diamond lattice method (see text). The dashed grey line indicates the diamond bond length. \textbf{a}, Fit result if spin 1 is fixed at the origin. \textbf{b}, Fit result if spin 14 is fixed at the origin.}
\label{fig:fitresult}
\end{figure}

The uncertainty of the fit for a given spin is dependent on its distance to the origin, which is set by the spin whose coordinates are fixed. This can hide information about the internal structure of a sub-cluster of strongly coupled spins that is far from the origin. In particular, the internal structure of sub-clusters might be tightly defined, while the position of the subcluster respectively to the origin is more uncertain. As an example, in Supplementary Fig. \ref{fig:fitresult}a, the origin is fixed at the position of spin 1, which is situated in a strongly coupled 8 spin sub-cluster. The uncertainties in the fitted positions for spins within this sub-cluster are therefore low, whereas the uncertainties in the fitted positions for spins 14 - 17, which are in a separate 4 spin sub-cluster, are higher. In Supplementary Fig. \ref{fig:fitresult}b, the origin is instead fixed at the position of spin 14, and consequently the uncertainties for the spins in the 4 spin sub-cluster are reduced, while for the 8 spin sub-cluster they are increased.

\subsection{Positioning the NV centre}\label{sec:NV_center_location}

The method presented in this work allows the structure of the measured nuclear spins to be obtained, without knowledge of the position of the NV relative to the nuclear spin cluster. In this section, we additionally determine the NV position. Using the coupling between the electron and nuclear spins to find the NV position is not straightforward, as the point-dipole approximation cannot be used due to the (unknown) electron wavefunction spread through the diamond lattice. Density functional theory (DFT) calculations \cite{Nizovtsev_2018} indicate that for spins with electron-nuclear couplings similar to those measured in this work, there can be large discrepancies from point-dipole couplings. 

\begin{figure}
\centering

\includegraphics[width=0.9\textwidth]{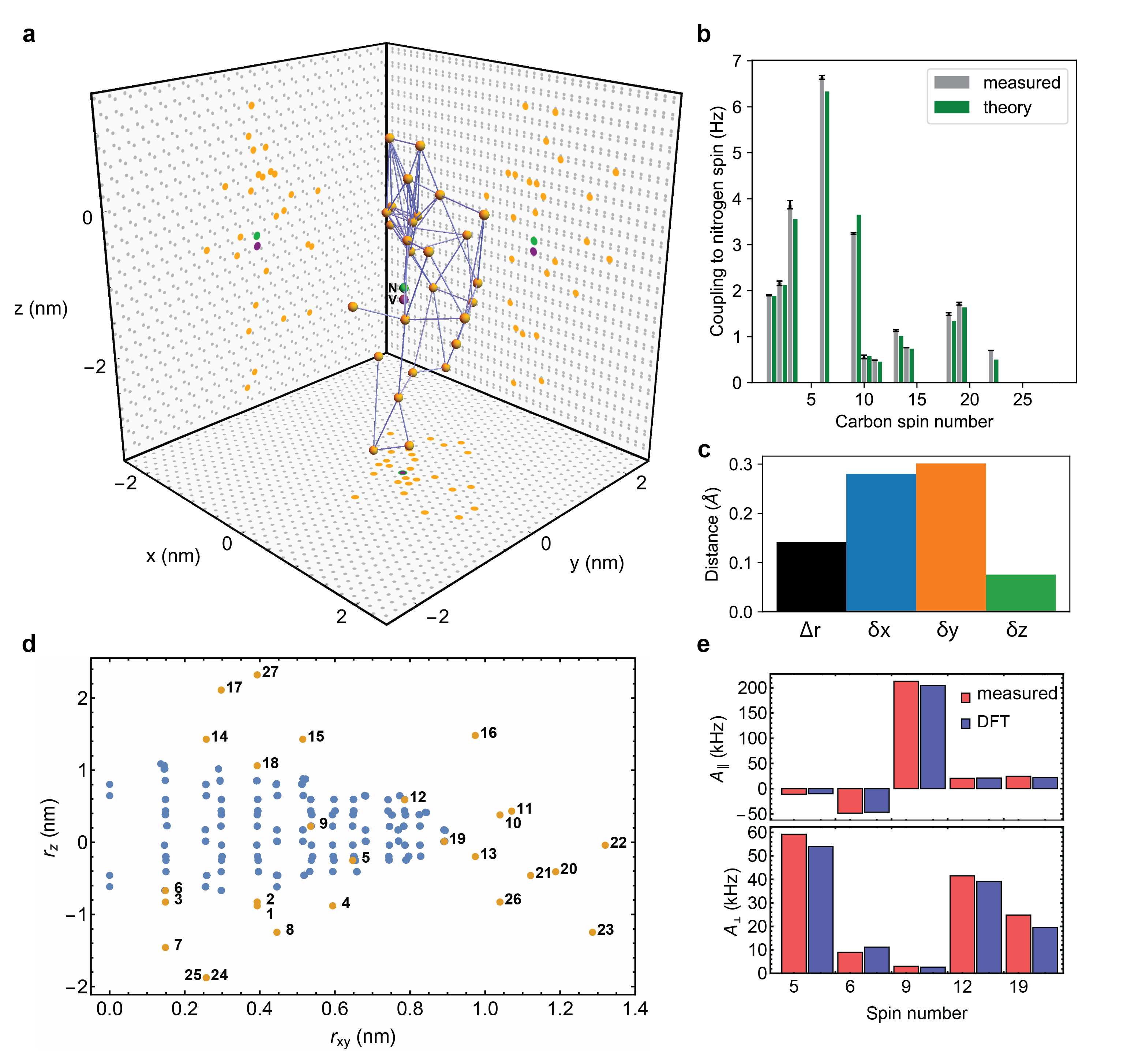}

\caption{{\bf Finding the position of the NV centre.} \textbf{a,} 3D plot showing the 27 spin cluster as shown in Supplementary Fig. \ref{fig:positioning}, along with the best solution for the nitrogen spin (green) and vacancy (purple) lattice sites calculated from the measured nitrogen-$^{13}$C couplings. \textbf{b,} Bar plot showing the measured couplings $f_{iN}$ between $^{13}$C spin $i$ and the nitrogen spin (grey), as well as the theoretically calculated couplings $|C_{iN}|/4\pi$ for the best solution (green). \textbf{c,} Bar plots of $\Delta r$ for the fitted solution for the nitrogen spin (black), as well as fit errors $\delta x$ (blue), $\delta y$ (orange) and $\delta z$ (green), where the $^{13}$C spins are fixed at the diamond lattice solution. \textbf{d,} Plot of $r_z$ vs. $r_{xy} = \sqrt{r_x^2 + r_y^2}$ for all lattice positions used in the DFT calculation from Nizovtsev et al. \cite{Nizovtsev_2018} (blue) and for the appropriately transformed $^{13}$C coordinates found in this work (orange, see text). Spins 5, 6, 9, 12 and 19 are close to a DFT lattice position, while the rest of the spins identified are outside of the 510 lattice sites simulated. \textbf{e,} Measured electron-$^{13}$C parallel (top) and perpendicular (bottom) hyperfine couplings for the 5 spins that are within the DFT calculation volume (red, taken from Supplementary Table \ref{table:carbonfrequencies}), compared with the DFT results (blue).}
\label{fig:NVpositioning}
\end{figure}

Our approach is to measure the couplings between the nitrogen nuclear spin and the $^{13}$C spins, as here the point-dipole approximation can be made. Nitrogen-$^{13}$C couplings can be measured using a similar double resonance procedure as for measuring $^{13}$C-$^{13}$C couplings. In this case, the nitrogen spin is first initialised in $\ket{m_I = 0}$ using measurement-based initialisation \cite{Robledo_Nature2011}, followed by a $\pi/2$ pulse resonant with the $\ket{m_s = \pm 1, m_I = 0} \leftrightarrow \ket{m_s = \pm 1, m_I = -1}$ transition. Using the nitrogen spin as the probe spin gives greater spectral resolution, as it has a longer $T_2^\text{echo}$ decay time (for the nitrogen $T_2^\text{echo}$ was measured to be $2.3(2)\,$s \cite{Bradley_arXiv2019}, whereas typically for $^{13}$C spins we measure $T_2^\text{echo} \sim 0.5\,$s). Supplementary Table \ref{table:couplingav} gives a number of measured couplings between the nitrogen and $^{13}$C spins, as shown in Supplementary Fig. \ref{fig:NVpositioning}b. Using the measured couplings, the nitrogen spin can be added to the $^{13}$C nuclear spin cluster using the method described in section \ref{sec:diamondlattice}, where $\gamma_j \rightarrow \gamma_n = 2\pi \times 0.3077\,$kHz/G, the nitrogen gyromagnetic ratio, in equation \ref{eq:hyperfine}. Determining the nitrogen lattice site also allows the vacancy site to be determined due to the known N-V distance and the alignment with the magnetic field along $z$, thereby giving the orientation of the NV centre with respect to the $^{13}$C nuclear spin cluster. The resulting 3D plot showing the best solution is shown in Supplementary Fig. \ref{fig:NVpositioning}a. The nitrogen spin coordinate is the same for all 5000 configurations identified. We can also use this position as an initial guess for a fit. Supplementary Fig. \ref{fig:NVpositioning}c shows the resulting difference between the initial guess and the fitted position, along with the associated fit uncertainties. 

Now that we independently determined the position of the $^{13}$C spins relative to the NV centre, we can compare the hyperfine couplings to DFT calculations, without any prior assumptions. In Nizovtsev et al. \cite{Nizovtsev_2018}, hyperfine couplings are calculated for 510 lattice sites surrounding the NV. Supplementary Fig. \ref{fig:NVpositioning}d shows a plot of lattice positions given in Nizovtsev et al. along with the coordinates of the $^{13}$C spins found in this work. The $^{13}$C spin coordinates are transformed so that the nitrogen spin is at the origin, and mirrored such that $z \rightarrow -z$, in order to be in the correct coordinate frame. Additionally a scaling factor of 1.02 was applied to correct for a different diamond lattice constant used, which was found by comparing the 510 lattice sites from Nizovtsev et al. with the same sites in our work. It can be seen that 5 of the 27 spins identified in this work overlap with lattice coordinates from Nizovtsev et al. The remaining spins lie outside of this region and therefore cannot yet be compared with DFT calculations. Supplementary Fig. \ref{fig:NVpositioning}e shows the measured electron-$^{13}$C hyperfine couplings (see Supplementary Table \ref{table:carbonfrequencies}), as well as those predicted in Nizovtsev et al., for the 5 spins. For the DFT results, we take the average of the predicted couplings for the possible C$_\text{3v}$ symmetric lattice sites. Additionally, we give the negative of the predicted $A_\parallel$ for all spins, as this gives better agreement with the measured results and a global minus sign is possible due to the unknown orientation of the magnetic field along $z$. It can be seen that the results give reasonably good agreement with DFT predictions, thereby giving an independent corroboration of the nuclear spin structure found in this work.

\subsection{Positioning spins using a cubic lattice}\label{sec:cubiclattice}

The method outlined in section \ref{sec:diamondlattice} is well suited for obtaining the structure of a $^{13}$C nuclear spin cluster in diamond. However, this method could not be used to find the structure of a spin cluster that is placed on the surface of the diamond, as the spins in this case are not confined to lie at diamond lattice points (although other a priori knowledge might be exploited). It is desirable to develop a method to find the structure of an arbitrary nuclear spin cluster. One option is to follow a similar method, but instead use a cubic lattice, which uses no pre-knowledge of the spin structure. In this case, the lattice spacing should be made much finer than in the diamond lattice case, so as to appropriately sample the entire volume given by the hyperfine coupling to the previous spin. The number of possible solutions per spin is consequently much larger, making the problem more computationally intensive. By using a cutoff for the number of solutions as introduced in section \ref{sec:diamondlattice}, the computation time can be reduced. However, this gives a risk that the optimum solution for the entire spin cluster will be discarded if a different solution is optimum for a subset of spins. 

Despite these computational challenges, it was found that a similar solution to the one found in section \ref{sec:diamondlattice} is obtained using this method. The cubic lattice was created with $2N_L + 1$ lattice points per edge of length $L$, where $N_L$ is an integer. This gives $(2N_L + 1)^3$ points in total in a volume of $L^3$, with lattice spacing $\Delta L = L/2N_L$. To further constrain the number of solutions obtained for each spin, $N_L$ and $L$ were varied depending on the coupling being used. By inverting equation \ref{eq:hyperfine}, we can find the maximum distance between two spins for a given coupling to be

\begin{equation}
    \Delta r_{ij}^\text{max} = \left( \frac{2\alpha_{ij}}{C_{ij}} \right)^{1/3},
\end{equation}
from which we can set $L_{ij} = 2\Delta r_{ij}^\text{max}$, with spin $i$ at the origin (see Supplementary Fig. \ref{fig:cubiclattice}a). We then set $N_L^{ij} = \tilde{N}_L/\Delta r_{ij}^\text{max}$, where $\tilde{N}_L = 2 \times 10^{-8}$ is a scaling factor chosen to give a fine enough lattice while also keeping the computation time within reasonable limits. Due to the rotational symmetry in the $x-y$ plane and inversion symmetry in $z$, for the second spin added the position is confined to the $x-z$ plane and for the third spin only solutions with positive $y$ values are taken.

Supplementary Fig. \ref{fig:cubiclattice}b-e show the result of this method. The routine took approximately 14 hours on a desktop PC parallelised over 8 cores. It can be seen that the best solution obtained is close to the diamond lattice solution, both in the predicted couplings (b,c) and the coordinates of the spins (d,e). The average distance between the diamond lattice positions and the cubic lattice positions is $0.58\,$\AA. We can also use this solution as an initial guess for least-squares minimisation, which returns a similar solution as that obtained by using the diamond lattice configuration as an initial guess (Supplementary Fig. \ref{fig:cubiclattice}e).

\begin{figure}
\centering

\includegraphics[width=0.9\textwidth]{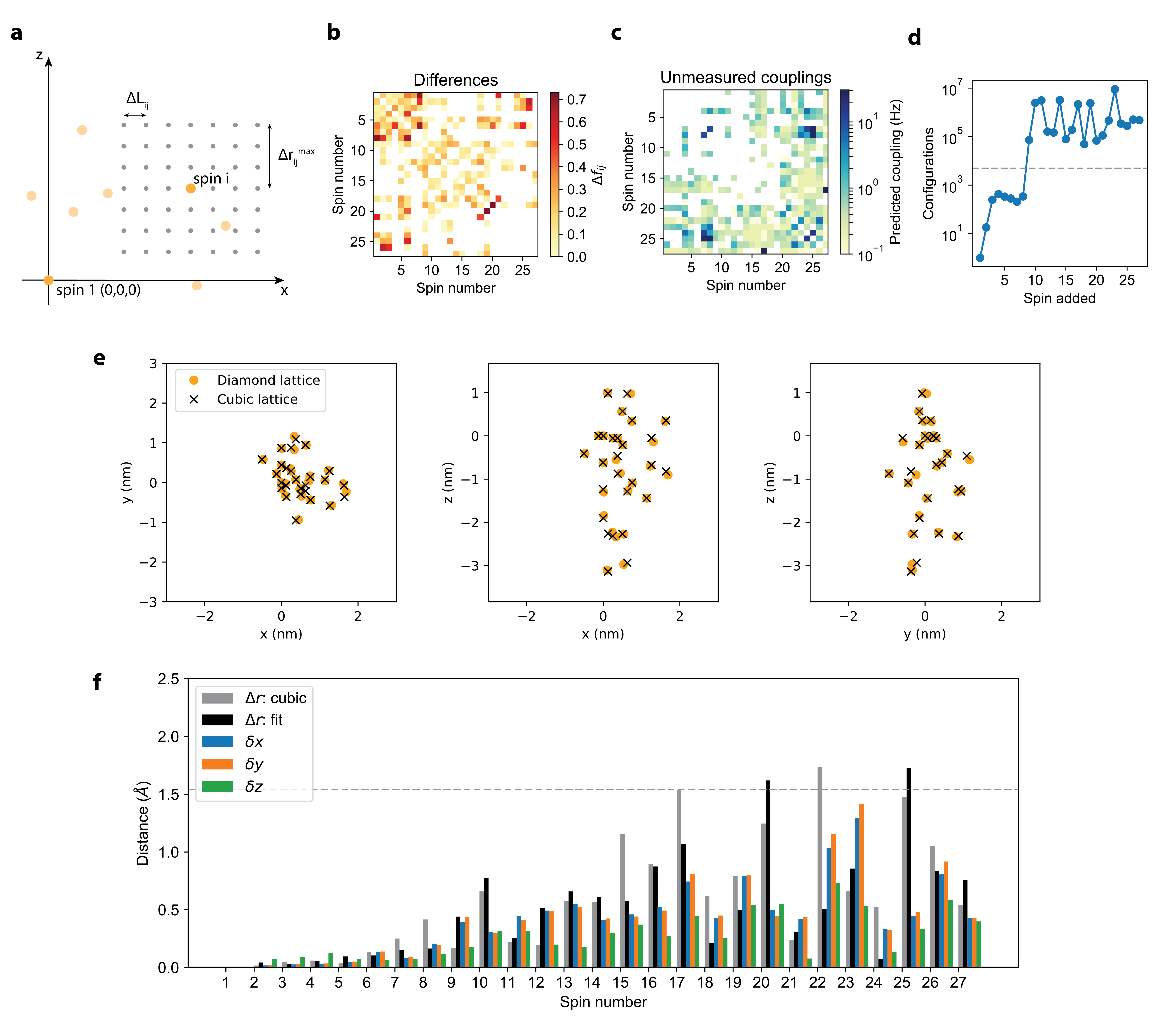}

\caption{{\bf Obtaining the structure using a cubic lattice.} \textbf{a,} Illustration of the lattice generation method, shown in 2D for clarity. Parameters are defined in the text. \textbf{b,} Matrix plot showing the residuals $\Delta f_{ij}$ for the configuration with the lowest $\xi$. White squares indicate unmeasured couplings. \textbf{c,} Matrix plot showing the predicted values of $|C_{ij}|/4\pi$ for couplings that were not measured. White squares indicate measured couplings. \textbf{d,} Total number of configurations found after each spin added (log scale). The dashed grey line indicates the cutoff point set at $X_\text{cutoff} = 5200$. \textbf{e,} 2D projections showing a comparison between the nuclear spin structure obtained using the diamond and cubic lattice methods. The cubic lattice solution has undergone the transformation $y \rightarrow -y$, $z \rightarrow -z$ plus a rotation around $z$ by $49.1\,$deg in order to align the two solutions. \textbf{f,} Distance between the diamond and cubic lattice solutions for each spin (grey), and fit result using the cubic lattice as an initial guess with spin 1 fixed, relative to the diamond solution (black). The associated fit errors $\delta x$ (blue), $\delta y$ (orange) and $\delta z$ (green) are also shown. The average distance between the cubic and diamond solutions is $0.58\,$\AA. }
\label{fig:cubiclattice}
\end{figure}

\clearpage

\bibliographystyle{naturemag}
\bibliography{Supp_bib}